\documentclass[fleqn,usenatbib]{mnras}

\usepackage{newtxtext,newtxmath}

\usepackage[T1]{fontenc}

\DeclareRobustCommand{\VAN}[3]{#2}
\let\VANthebibliography\thebibliography
\def\thebibliography{\DeclareRobustCommand{\VAN}[3]{##3}\VANthebibliography}

\usepackage{graphicx}	
\usepackage{amsmath}	

\renewcommand{\vec}[1]{\ensuremath{\boldsymbol{#1}}}
\newcommand{\nvec}[1]{\ensuremath{\boldsymbol{\hat{#1}}}}
\newcommand{\minone}{$^{-1}$}
\newcommand{\HL}[1]{\textcolor{black}{#1}}

\title[Introducing \texttt{cuDisc}: a 2D PPD code]{Introducing \texttt{cuDisc}: a 2D code for protoplanetary disc structure and evolution calculations}

\author[A. Robinson et al.]{
Alfie Robinson,$^{1}$\thanks{E-mail: a.robinson21@imperial.ac.uk}
Richard A. Booth$^{2}$ and
James E. Owen$^{1}$
\\
$^{1}$Astrophysics Group, Imperial College London, Prince Consort Road, London SW7 2AZ, UK\\
$^{2}$School of Physics and Astronomy, University of Leeds, Leeds, LS2 9JT, UK
}

\date{Accepted XXX. Received YYY; in original form ZZZ}

\pubyear{2015}

\begin{document}
\label{firstpage}
\pagerange{\pageref{firstpage}--\pageref{lastpage}}
\maketitle

\begin{abstract}
We present a new 2D axisymmetric code, \texttt{cuDisc}, for studying protoplanetary discs, focusing on the self-consistent calculation of dust dynamics, grain size distribution and disc temperature. Self-consistently studying these physical processes is essential for many disc problems, such as structure formation and dust removal, given that the processes heavily depend on one another. To follow the evolution over substantial fractions of the disc lifetime, \texttt{cuDisc} uses the \texttt{CUDA} language and libraries to speed up the code through GPU acceleration. \texttt{cuDisc} employs a second-order finite-volume Godonuv solver for dust dynamics, solves the Smoluchowski equation for dust growth and calculates radiative transfer using a multi-frequency hybrid ray-tracing/flux-limited-diffusion method. We benchmark our code against current state-of-the-art codes. Through studying steady-state problems, we find that including 2D structure reveals that when collisions are important, the dust vertical structure appears to reach a diffusion-settling-coagulation equilibrium that can differ substantially from standard models that ignore coagulation. For low fragmentation velocities, we find an enhancement of intermediate-sized dust grains at heights of $\sim 1$ gas scale height due to the variation in collision rates with height, and for large fragmentation velocities, we find an enhancement of small grains around the disc mid-plane due to collisional ``sweeping'' of small grains by large grains. These results could be important for the analysis of disc SEDs or scattered light images, given these observables are sensitive to the vertical grain distribution.
\end{abstract}




\section{Introduction}

The past few decades have seen the study of young planetary systems and their formation environments become a rapidly evolving field with major interest within the astrophysics community. This is largely due to an unprecedented wealth of observations made possible by recent observatories such as ALMA \citep{alma2009} and the continual discovery of diverse exoplanetary systems \cite[e.g][]{mayor2011,kepler2013,winn15,madhusudhan19,zhu21}. Protoplanetary discs, the discs of gas and dust that form around young stars, are the birthplace of such planetary systems and have, therefore, been subject to extensive theoretical interest. Various questions relating to their nature have arisen in recent years that remain at least partly unanswered by current theoretical models. Examples of such problems include: the nature of the mechanisms behind ``sub-structure'' formation in protoplanetary discs, as observations have shown these objects to exhibit diverse features from axisymmetric rings and gaps to non-axisymmetric arcs \citep{andrews20,bae22}; the connection between the spatial distribution and evolution of chemical species in discs to the eventual compositions of planetary cores and atmospheres \citep{oberg2011,booth17,madhusudhan19, eistrup23}; and mechanisms for the dispersal of protoplanetary discs after their observed lifetimes of $\sim$ a few Myr \citep{ercolano17,owen2019}. \\

Exploring these problems requires sophisticated numerical modelling, given the plethora of physical processes that govern the structure and evolution of protoplanetary discs. Our understanding of discs has primarily been advanced through the use of state-of-the-art codes for studying the dynamics and thermodynamics of the gas and dust that comprise the disc material. 2D and 3D simulations have typically been used to study discs on short, dynamical time-scales, given their computational cost, whilst 1D models have often been used to study discs on longer, secular time-scales. Work done using these models has hugely advanced our understanding of protoplanetary discs; however, it has become evident that for certain problems, the interplay of each of the facets of disc physics - dynamics, thermodynamics and the dust size distribution - must be studied self-consistently over secular time-scales. The 1D code \texttt{DustPy} \citep{Stammler_2022} is the current state-of-the-art for studying problems of this nature; however, it cannot be used if the problem depends on the intricacies of the disc vertical structure. Examples of such problems include temperature instabilities \citep{watanabe2008, wu2021, David_Melon_Fuksman_2022} where 2D temperature solvers have been used but dust dynamics and growth neglected, snow line instabilities \citep{owen2020} where 1D temperature and dynamics solvers have been used, and problems relating to disc dispersal such as the removal of dust from discs via radiation pressure \citep{owen2019, krum2020}, which has been studied with both 1D and 2D solvers but without dust growth/fragmentation, or entrainment in photoevaporative or magnetic winds \citep{franz2020, booth21, hutch2021, roden2022}, again with 1D and 2D approaches but with dust grain growth neglected. \\

This paper presents a new code, \texttt{cuDisc}, that includes 2D multi-fluid dust dynamics coupled to both 2D temperature evolution and 1D secular gas evolution. Dust and gas can be evolved, whilst simultaneously evolving the dust grain size distribution and the resulting temperature structure. This self-consistent calculation means fewer assumptions about the system's state for a particular scenario must be made. The code uses a 2D grid in the poloidal plane, meaning that assumptions about vertical structure in the dust do not have to be made as the structure can be calculated self-consistently. In order to allow evolution calculations to be run for significant fractions of the typical disc lifetime ($\sim$ Myr) in computationally feasible time-scales, \texttt{cuDisc} utilises GPU acceleration via the employment of the CUDA language and libraries. In the rest of this paper, Sections 2, 3, 4 and 5 outline the numerical methods used by \texttt{cuDisc} whilst Section 6 outlines the code structure and Section 7 shows some example science calculations made using \texttt{cuDisc} where we study grain growth in a steady-state transition disc.

\section{Grid Structure}

\begin{figure}
    \centering
    \includegraphics[width=0.9\columnwidth,trim={0cm 2.5cm 4cm 7cm}, clip]{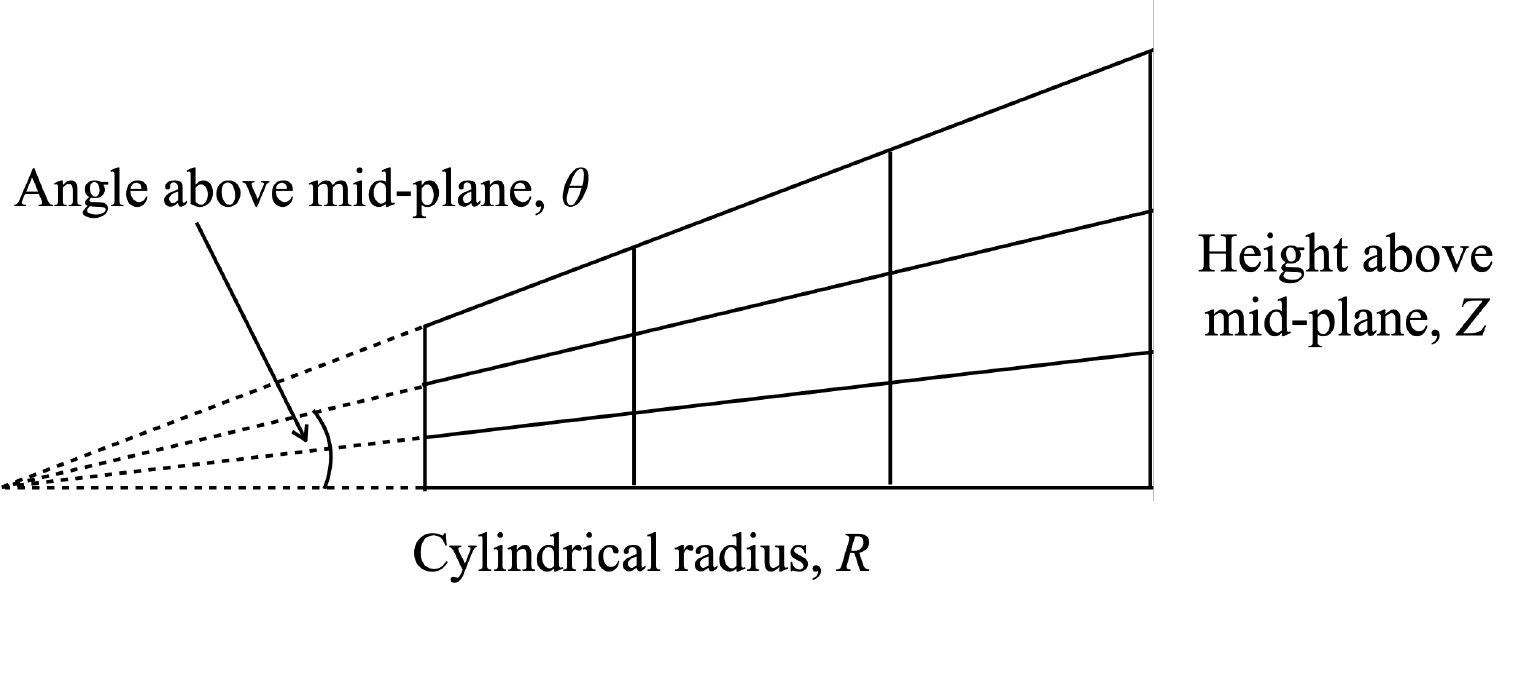}
    \caption{The computational grid used in \texttt{cuDisc}. Cell boundaries exist along lines of constant radius and constant angle above the mid-plane, $\theta$.}
    \label{grid}
\end{figure}

\begin{figure}
    \centering
    \includegraphics[width=0.85\columnwidth,trim={2cm 0cm 0cm 0cm}, clip]{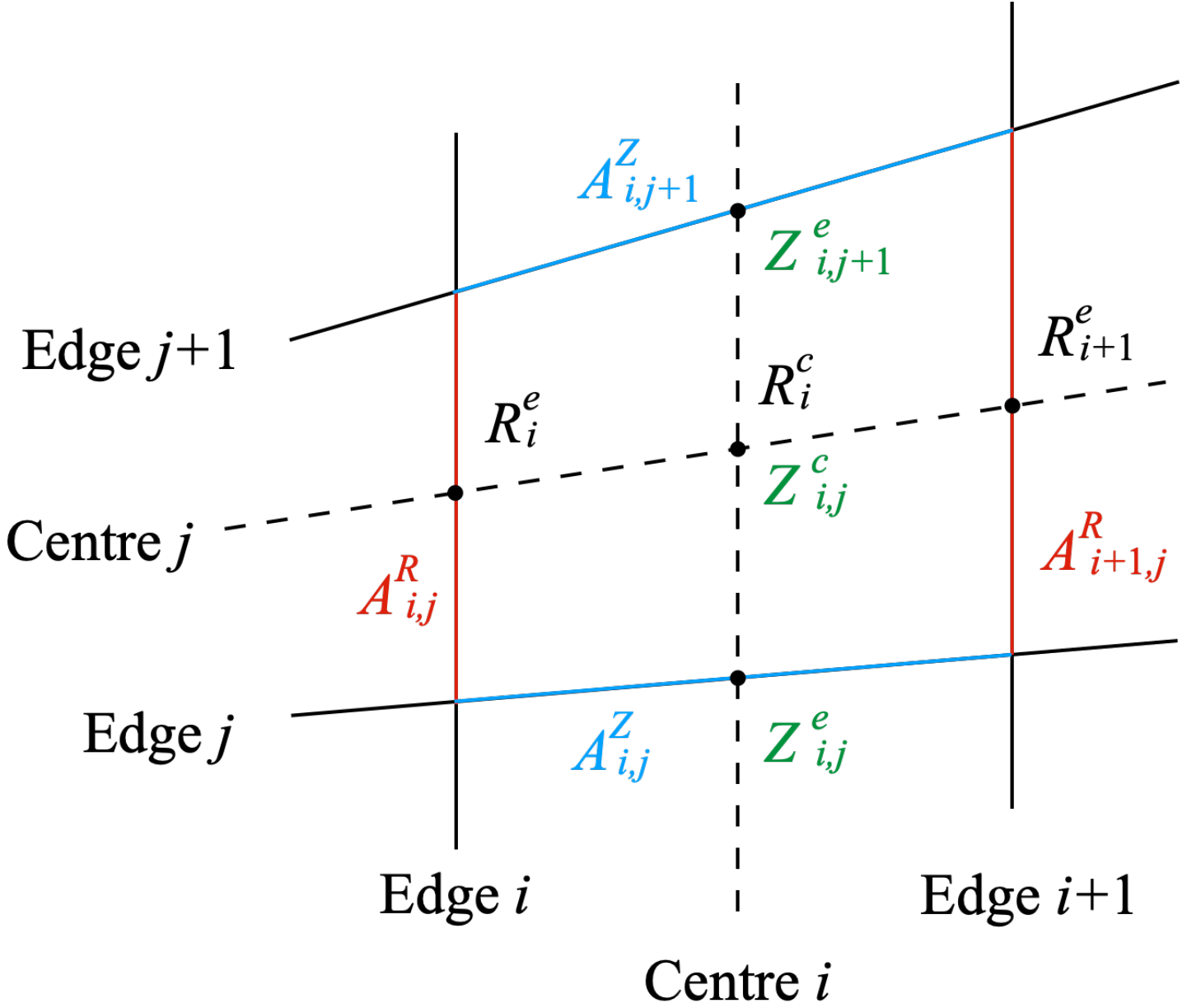}
    \caption{The cell structure and indexing implemented in \texttt{cuDisc}. The indices in the $R$ and $Z$ directions are $i$ and $j$, respectively.}
    \label{cell}
\end{figure}

\texttt{cuDisc} is a 2D code in the poloidal plane, with axisymmetry adopted about the disc's rotation axis. The grid is structured in a fashion that makes certain features of disc physics easier to calculate; cell interfaces are defined along lines of constant polar angle above the mid-plane, $\theta$, and constant cylindrical radius, $R$, as shown in Fig. \ref{grid}. This means that the basis vectors of the grid coordinate system are spherical radius, $\nvec{r}$, and cylindrical height, $\nvec{Z}$, although vector quantities (such as velocity) are dealt with in the standard cylindrical components. This grid structure allows for ray-tracing from the central star for calculating quantities such as optical depth whilst also making calculations that require vertical integration easy, such as computing hydrostatic equilibrium. Cell spacing can be arbitrary in principle, but the standard implementation is logarithmic in $R$ with a power law in $\theta$. The individual cell structure is shown in Fig. \ref{cell}. Physical quantities are stored at cell centres, and edge coordinates are required for the reconstruction of the cell-centred quantities to the cell interfaces for the advection routine (see later sections). Cell volumes, $V_{i,j}$, and interface areas $A^R_{i,j}$ \& $A^Z_{i,j}$ are given by 
\begin{equation}
    V_{i,j} = \frac{1}{3} d(R^3)_{i} d(\tan\theta)_{j},
\end{equation}
\begin{equation}
    A^R_{i,j} = (R^e_{i})^2  d(\tan\theta)_{j},
\end{equation}
\begin{equation}
    A^Z_{i,j} = \frac{1}{2} \frac{d(R^2)_{i}}{\cos\theta^e_j}.
\end{equation}

These geometric factors are written in such a way as to avoid divergence as $R \rightarrow 0$. Explicitly, $d(R^n)_{i} = (R^e_{i+1})^n - (R^e_{i})^n$ and $d(\tan\theta)_{j} = \tan\theta^e_{j+1} - \tan\theta^e_{j}$. The $Z$ coordinates are then given by $Z^e_{i,j}=R^c_i \tan\theta^e_{i,j}$ and $Z^c_{i,j}=R^c_i \tan\theta^c_{i,j}$. Due to the non-orthogonality of the grid coordinate bases, when calculating fluxes through interfaces, the dot-product of velocity with the interface normals must be calculated to only account for the velocity component orthogonal to the interface. \\

At each edge of the active cell domain, ghost cells are employed in order to set the boundary conditions. The conditions along each of the four boundaries can be set independently as either: outflow, where ghost cell quantities are set by the values in the first adjacent active cell, zero, where ghost cell quantities are set to floor values, or closed, where ghost cell quantities are set by the values in the adjacent active cell but with zero flux over the boundary. The default setup assumes vertical symmetry in the disc about the mid-plane and, therefore, uses a minimum $\theta$ of 0 with a closed boundary condition and outflow boundary conditions for the three other boundaries.
\section{Dynamics Solver}

In \texttt{cuDisc}, the dust species are evolved by treating each grain size as a pressureless fluid and solving their associated advection-diffusion equations. We employ a second-order finite-volume Godonuv scheme for solving the set of equations \citep{STONE2009139}. We write our equations in terms of vector fields of the conserved quantities, $\vec{Q}$, their associated fluxes, $\vec{F}(\vec{Q})$, and any source terms, $\vec{S}$. For our system, these fields are given by
\begin{equation}
    \label{quants}
    \vec{Q}_i = \begin{pmatrix} \rho_i \\ \rho_i v_{R,i} \\ \rho_i v_{\phi,i} R \\ \rho_i v_{Z,i} \end{pmatrix}, 
\end{equation}
where $\rho_i$ and $\vec v_i = (v_{R,i}, v_{\phi,i}, v_{Z,i})$ are the volume density and velocity of dust species $i$ respectively,
\begin{equation}
    \label{dynfluxes}
     \vec F_i = \begin{pmatrix} \rho_i \vec v_i + \vec F_{\text{diff},i} \\ v_{R,i} (\rho_i \vec v_i + \vec F_{\text{diff},i})\\ v_{\phi,i} R (\rho_i \vec v_i + \vec F_{\text{diff},i})\\ v_{Z,i} (\rho_i \vec v_i + \vec F_{\text{diff},i}) \end{pmatrix},
\end{equation}
where the two terms for each conserved quantity are the fluxes generated by advection, $\rho_i \vec v_i$, and diffusion, $\vec F_{\text{diff},i}$,
\begin{equation}
    \label{sources}
    \vec S_i =  \begin{pmatrix} 0 \\ \dfrac{\rho_i v_{\phi,i}^2}{R} - \rho_i \Omega^2 R + f_{\text{drag},i,R} + f_{\text{ext},i,R} \vspace{5pt} \\  f_{\text{drag},i,\phi} \vspace{5pt} \\ - \rho_i \Omega^2 Z + f_{\text{drag},i,Z} + f_{\text{ext},i,Z} \end{pmatrix},
\end{equation}
where $\Omega$ is the Keplerian angular velocity, $\sqrt{GM_* / (R^2+Z^2)^{3/2}}$, and $\vec f_{\text{drag}}$ and $\vec f_{\text{ext}}$ are source terms that arise due to dust-gas drag and any external forces (e.g. radiation pressure). We formulate diffusion in the momentum equations as the diffusive flux acting to diffuse the advective quantities. We do not currently consider the other terms discussed in recent works, such as \cite{huang2022}\HL{; these being the advection of the diffusive flux and the time-dependent diffusive flux. Diffusion related terms arise out of modelling turbulence by writing the density and velocity components as a short-term average plus a short-term fluctuation, then averaging the resulting equations over the short-term, keeping any terms that include correlations between fluctuations (Reynolds averaging, see e.g. \citealt{cuzzi1993}).} Note that we solve an azimuthal equation even though all $\phi$ derivatives are zero by axisymmetry, as the advection of angular momentum throughout the $(R,Z)$ plane could be important in some problems; this is sometimes referred to as 2.5D. Also, note that we write this equation in angular momentum conserving form as this removes the Coriolis force term, which has been shown to lead to loss of angular momentum conservation \cite[see e.g.][]{kley1998}. \\

The quantities, $\vec Q$, are updated by solving the advection-diffusion equations given by 
\begin{equation}
    \label{advgen}
    \frac{\partial \vec{Q}}{\partial t} + \nabla \cdot \vec{F} = \vec S.
\end{equation}
We solve this set of equations in two stages via operator splitting: a transport step where the advection-diffusion equations are solved as homogeneous hyperbolic equations and a source step where the quantities are updated through the source terms. For the transport step, the equations are integrated over volume to find the flux-conservative form. Discretising this gives the equation used to evolve a quantity $Q$ in cell $i,j$ to time $n+1$,
\begin{equation}
    \label{advdisc}
    Q^{n+1}_{i,j} = Q^n_{i,j} - \frac{\Delta t}{V_{i,j}} (FA)^{n+\frac{1}{2}}_{i,j},
\end{equation}
where $(FA)_{i,j}$ is the total area-weighted-flux exiting through the interfaces between cell $i,j$ and adjacent cells, given by
\begin{multline}
    \label{fluxarea}
    (FA)_{i,j} = F^R_{i+1,j} A^R_{i+1,j} - F^R_{i,j} A^R_{i,j} \\
    + F^Z_{i,j+1} A^Z_{i,j+1} - F^Z_{i,j} A^Z_{i,j}.
\end{multline}
The time index of $n+\frac{1}{2}$ indicates that the fluxes should be calculated at the half time-step to represent the average flux over the full time-step $\Delta t$. Following \cite{STONE2009139}, we calculate these half-time fluxes in the following manner:

\begin{enumerate}
    \item The primitive quantities, $(\rho_i, \vec v_i)$, are reconstructed at both sides of the cell interfaces from the conserved quantities at cell centres using a first-order donor cell method and a Riemann solver is used to calculate first-order advective fluxes through each interface at time $n$.
    \item The diffusive fluxes at each interface at time $n$ are calculated from the cell-centred conserved quantities and added to the advective fluxes to get the total area-weighted flux exiting each cell at time $n$, $(FA)^n_{i,j}$.
    \item The conserved quantities are advanced by a half time-step using these fluxes:
    \begin{equation}
        \label{halftimestepupdate}
        Q^{n+\frac{1}{2}}_{i,j} = Q^n_{i,j} - \frac{1}{2} \frac{\Delta t}{V_{i,j}} (FA)^{n}_{i,j},
    \end{equation}
    \item These half-updated quantities are given a half time-step source update (described in detail in \autoref{ssec:source}).
    \item Primitive quantities at time $n+\frac{1}{2}$ are now reconstructed from these half-updated quantities using a second-order piecewise linear method with a modified van Leer limiter detailed by \cite{mignone2014}. These half-time primitive quantities are used to calculate second-order advective fluxes through each interface at time $n+\frac{1}{2}$.
    \item The diffusive fluxes at each interface at time $n+\frac{1}{2}$ are calculated from the half-updated cell-centred quantities and added to the second-order advective fluxes to get the total area-weighted flux exiting each cell at time $n+\frac{1}{2}$, $(FA)^{n+\frac{1}{2}}_{i,j}$.
\end{enumerate}

In steps (i) and (v), where the advective fluxes are calculated, the Riemann problem has to be solved at the cell interfaces. For pressureless fluids, collisions between fluids with different velocities lead to ``delta shocks'', which travel at the Roe-averaged velocity \citep{ROE1981357},
\begin{equation}
    \label{roeav}
    \hat{v} = \frac{\sqrt{\rho_l}\vec v_l \cdot \nvec{n} + \sqrt{\rho_r} \vec v_r \cdot \nvec{n}}{\sqrt{\rho_l} + \sqrt{\rho_r}},
\end{equation}
where $r$ and $l$ denote the values of the reconstructed quantities to the left and right of the interface in question. The reconstructed velocities are dotted with the interface normal, $\nvec{n},$ to find the velocity orthogonal to each interface; this is important given that we use a non-orthogonal grid. Numerically, we capture this process according to \cite{leveque04} by taking the upwind flux depending on the sign of the Roe-averaged velocity:
\begin{equation}
    \label{riemann}
    F_\text{adv} = 
    \begin{cases}
        Q_l \vec v_l \cdot \nvec{n} & \text{if } \hat{v} > 0\\
        Q_r \vec v_r \cdot \nvec{n} & \text{if } \hat{v} < 0\\
        \frac{1}{2} (Q_l \vec v_l \cdot \nvec{n}+ Q_r \vec v_r \cdot \nvec{n}) & \text{if } \hat{v} = 0\\
        0 & \text{if }\vec v_l \cdot \nvec{n}< 0 < \vec v_r \cdot \nvec{n},  
    \end{cases}
\end{equation}
The diffusive fluxes are then added to these advective fluxes to find the total flux. 

\subsection{Dust source terms and diffusion}
\label{ssec:source}
The source terms given by Eqn. \ref{sources} include curvature terms, gravitational terms, drag terms and any other external forces. Drag is caused by dust-gas interactions and is given by
\begin{equation}
    \label{dragforce}
    \vec f_{\text{drag}} = -\frac{\rho}{t_{s}} (\vec v - \vec v_g),
\end{equation}
where $\vec v_g$ is the gas velocity and $t_{s}$ is the stopping time that conveys how well-coupled dust grains are to the gas flow; short stopping times imply well-coupled dust grains that are entrained in the gas flow. For the problems that we intend to study using \texttt{cuDisc}, the largest dust grains have radii $\sim$ cm size, meaning that we are in the regime of Epstein drag. This gives the stopping time as 
\begin{equation}
    \label{t_stop}
    t_{s} = \frac{\rho_{m} s}{\rho_g v_\text{th}},
\end{equation}
where $\rho_{m}$ is the dust grain internal density, $s$ is the dust grain radius, $\rho_g$ is the gas mass density and $v_\text{th}$ is the thermal velocity of the gas, usually given by $(8/\pi)^{1/2} c_s$, where $c_s$ is the isothermal gas sound speed.\\

Including source term updates, the full time-step is given by:

\begin{enumerate}
    \label{full_dt}
    \item $\vec Q^*_{i,j} =\vec Q^n_{i,j} - \dfrac{1}{2} \dfrac{\Delta t}{V_{i,j}} \vec{(FA)}^{n}_{i,j}$
    \item $\vec Q^{**}_{i,j} = \vec Q^*_{i,j} + \dfrac{1}{2}\Delta t \vec S_{\text{exp}}(\vec P^n_{i,j})$ 
    \item $\vec Q^{**}_{i,j} \rightarrow  \vec P^{**}_{i,j} $
    \item $\vec P^{n+\frac{1}{2}}_{i,j} = \left(\rho^{**}_{i,j}, \; \vec v^{**}_{i,j} - \dfrac{\Delta t /2}{\Delta t /2 + t_s} ( \vec v^{**}_{i,j} - \vec v_{g,i,j}) \right)$
    \item $\vec Q^{\dag}_{i,j} = \vec Q^n_{i,j} - \dfrac{\Delta t}{V_{i,j}} \vec{(FA)}^{n+\frac{1}{2}}_{i,j}$
    \item $\vec Q^{\dag \dag}_{i,j} = \vec Q^{\dag}_{i,j} + \Delta t \vec S_{\text{exp}}(\vec P^{n+\frac{1}{2}}_{i,j})$ 
    \item $\vec Q^{\dag \dag}_{i,j} \rightarrow  \vec P^{\dag \dag}_{i,j} $
    \item $\vec P^{n+1}_{i,j} = \left(\rho^{\dag \dag}_{i,j}, \; \vec v^{\dag \dag}_{i,j} - \dfrac{\Delta t}{\Delta t + t_s} ( \vec v^{\dag \dag}_{i,j} - \vec v_{g,i,j}) \right) $
\end{enumerate}
where $\vec Q$ \& $\vec P$ represent conserved quantities and primitive quantities respectively, $\vec S_{\text{exp}}$ is the vector of source terms that are solved explicitly (curvature, gravity and external forces) and quantities with an asterisk ($*$) or a dagger ($\dag$) signify intermediate states. Steps (iii) and (vii) signify conversions of conserved quantities to primitive quantities. Steps (iv) and (viii) are the drag updates to the primitive velocities; these are solved implicitly, as if they were solved explicitly, then short stopping times would prohibitively limit the length of an explicit time-step. As the implicit update only includes the drag terms, the Jacobian for the implicit system of equations is linear and diagonal, meaning the solution has a simple, exact analytic form. Boundary conditions are set before step (i) and again after step (iv).\\

The diffusive mass flux discussed in the previous section arises due to the diffusion of dust grains via turbulent gas motions. This diffusive velocity is calculated in a way analogous to molecular diffusion  \cite[][]{clarke1988,taklin02}, where the diffusive flux is given by
\begin{equation}
    \label{diffflux}
    \vec{F}_{\text{diff},i} = - \frac{\rho_g \nu}{\text{Sc}_{\HL{i}}} \nabla \left( \frac{\rho_i}{\rho_g} \right),
\end{equation}
where $\nu$ is the kinematic viscosity of the disc, and Sc\HL{$_i$} is the Schmidt number, a dimensionless number that represents the strength of the dust-gas coupling \HL{ for a given grain}. \HL{In the calculations presented in this paper, Sc$_i$ is set to unity for all dust species, however the code allows any choice for the form of Sc$_i$, which may vary arbitrarily with both position and dust properties.} Given that we formulate the diffusion in the momentum equations as the diffusive flux acting to diffuse the advective quantities, the diffusive momentum flux at an interface is given by $\vec{F}_{\text{diff},i}$ multiplied by the left or right advective velocity depending on the sign of the diffusive flux itself,
\begin{equation}
    \label{Fdiffmom}
    F_\text{diff,mom} = 
    \begin{cases}
        F_\text{diff} \vec v_l \cdot \nvec{n} & \text{if } F_\text{diff} > 0\\
        F_\text{diff} \vec v_r \cdot \nvec{n} & \text{otherwise.}\\ 
    \end{cases}
\end{equation}

These diffusive fluxes are then added to the advective fluxes defined earlier to give the total flux.\\

The time interval $\Delta t$ for each time-step is determined from the Courant-Friedrichs-Lewy (CFL; \citealt{Courant_1928}) condition:
\begin{equation}
    \label{dtCFL}
    \Delta t_\text{CFL} 
 = \text{min}\left( \Delta t^\text{adv}_\text{CFL}, \Delta t^\text{diff}_\text{CFL} \right),
\end{equation}
where $\Delta t^\text{adv}_\text{CFL}$ \& $\Delta t^\text{diff}_\text{CFL}$ are the CFL time-steps for advection and diffusion respectively, given by
\begin{equation}
    \label{dt_adv}
    \Delta t^\text{adv}_\text{CFL} = C^\text{adv} \text{min}\left(|dR_i/v^s_{R,i,j}|,|dZ_{i,j}/v^s_{Z,i,j}|\right),
\end{equation}
where the minimum is over all cells $i,j$ and species $s$, and 
\begin{equation}
    \label{dt_diff}
    \Delta t^\text{diff}_\text{CFL} = C^\text{diff} \text{min}\left(|(dR_i)^2/(\nu_{i,j}/\text{Sc}_s)|,|(dZ_{i,j})^2/(\nu_{i,j}/\text{Sc}_s)|\right).
\end{equation}
Both time-steps have associated Courant numbers, $C^\text{adv}$ \& $C^\text{diff}$, which are safety factors set by default to 0.4 and 0.2, respectively.
\subsection{Diffusion}\label{diffusion}

For both the dust dynamics and the radiative diffusion problem presented in Section \ref{tempsolv}, we need to calculate diffusive fluxes. In general, a diffusive flux can be written as
\begin{equation}
    \label{diffgen}
    \vec F = - D \nabla u,
\end{equation}
where $D$ is the diffusion constant and $u$ is the quantity being diffused. \\

Calculating the diffusive flux requires care in \texttt{cuDisc} given the non-orthogonal coordinate system in use. This is because using the standard difference formulae would give gradients of $u$ that are not perpendicular to the interface in question and therefore using them directly would provide incorrect fluxes across the interfaces. To deal with this we need a method that calculates the 2D gradient of $u$ that we can dot-product with the interface normal to find the flux across the interface. To this end, we use the second-order accurate, conservative method detailed by \cite{WU20127152} for diffusion problems on arbitrary mesh structures. Here, we present this method specifically for our mesh. \\

\begin{figure}
    \centering
    \includegraphics[width=0.95\columnwidth, trim={3cm 2cm 3cm 2cm}, clip]{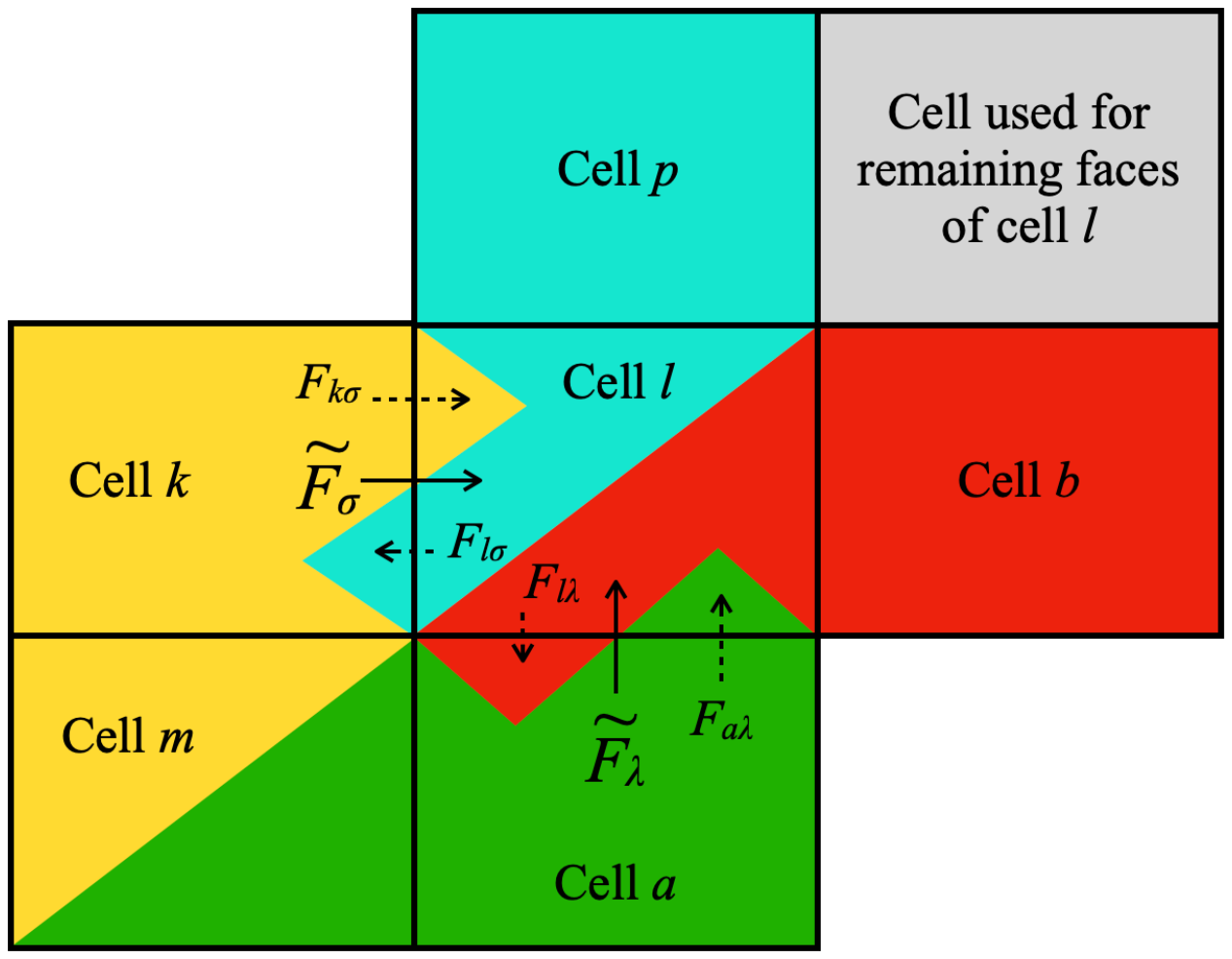}
    \caption{The cell stencil used for calculating diffusive fluxes over interfaces. The cells are shown as orthogonal for illustrative ease. The total flux $\Tilde{F}_\sigma$ through the interface $\sigma$ that connects cells $k$ and $l$ is calculated using the one-sided fluxes, $F_{k\sigma}$ and $F_{l\sigma}$. The same can be said for the total flux $\Tilde{F}_\lambda$ for interface $\lambda$ but with the one-sided fluxes $F_{a\lambda}$ and $F_{l\lambda}$. The cell colours indicate which cells are used to construct the one-sided fluxes: cells with yellow for $F_{k\sigma}$, cyan for $F_{l\sigma}$, green for $F_{k\lambda}$, and red for $F_{l\lambda}$.}
    \label{cellstencil}
\end{figure}

To compute the gradients of $u$ within a cell, Wu et al. introduce interpolation points on the cell interfaces. The location of these interpolation points and accompanying values of $u$ are determined through the conditions (i) $u$ must be continuous everywhere and (ii) the flux across the interface must be continuous, i.e. $D \nvec{n} \cdot \nabla u$ must be continuous at the interface, $\nvec{n}$ being the interface normal vector. In this paper, we describe how these interpolation points are found; for readers interested in why these criteria lead to the interpolation points, we refer to the appendix of \cite{WU20127152}. The flux at an interface is built up from one-sided fluxes on each side of an interface. Take a radial interface $\sigma$ between cells $k$ and $l$ (see Fig. \ref{cellstencil}). The net flux across the face from the perspective of cell $k$ is given by 
\begin{equation}
    \label{fluxes}
    \Tilde{F}_\sigma = w_{k\sigma} F_{k\sigma} - w_{l\sigma} F_{l\sigma},
\end{equation}
where $w_{k\sigma}$ and $w_{l\sigma}$ are weights assigned to the one-sided flux contributions from the cells on each side of the interface, $F_{k\sigma}$ and $F_{l\sigma}$, where these fluxes are directed out of cells $k/l$ across the interface $\sigma$. The weights are defined as
\begin{equation}
    \label{weights}
    w_{k\sigma} = \frac{\mu_{k\sigma}}{\mu_{k\sigma} + \mu_{l\sigma}}, \hspace{10pt} w_{k\sigma} + w_{l\sigma} = 1,
\end{equation}
where $\mu_{k\sigma} = \frac{d_{k\sigma}}{D_k}$, $d_{k\sigma}$ being the shortest (perpendicular) distance between the cell centre and the plane of the interface $\sigma$ and $D_k$ the diffusion constant associated with cell $k$. \\

\begin{figure}
    \centering
    \includegraphics[width=0.9\columnwidth,trim={2cm 2cm 2cm 2cm}, clip]{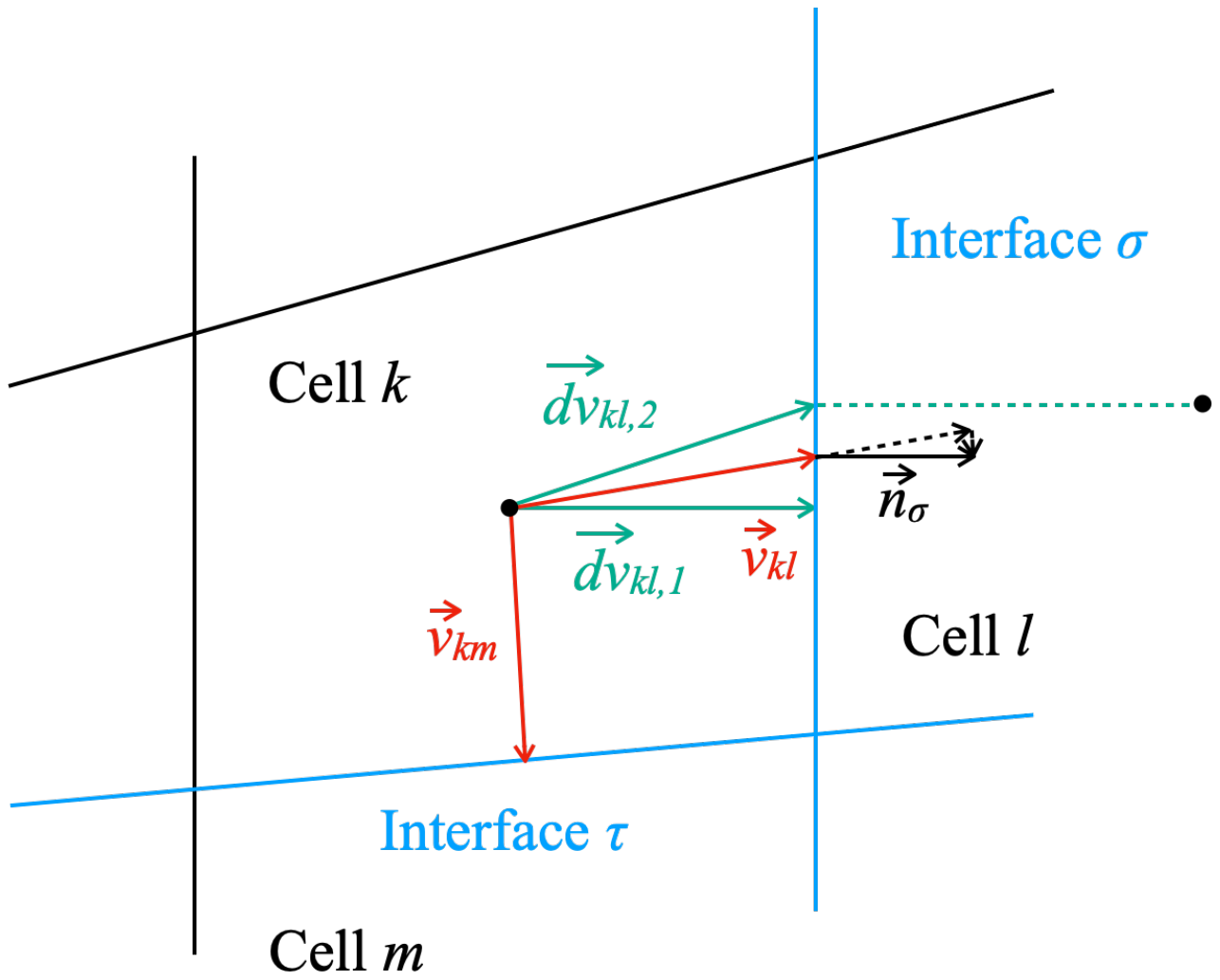}
    \caption{The vector decomposition used for the cell interface normals when calculating diffusive fluxes. Green vectors show those that point from cell centre $k$ to the points on the interface $\sigma$ that are closest to cell centres $k$ and $l$. The red vector $\vec v_{kl}$ is then constructed via a weighted sum of the two green vectors. $\vec v_{km}$ is constructed from vectors for the interface between cells $k$ and $m$ but are not shown for simplicity.}
    \label{vecdecomp}
\end{figure}

To include all flux orthogonal to the interface $\sigma$, the one-sided flux $F_{k\sigma}$  is calculated by decomposing the interface normal vector $\vec n_\sigma$ into two vectors, $\vec v_{kl}$ and $\vec v_{km}$, that point from cell centre $k$ to the interpolation points on the cell interface in question and the clockwise-adjacent interface that joins cell $k$ to cell $m$, as shown in Fig. \ref{vecdecomp}. To find the interpolation point $\vec v_{kl}$ we first find two vectors that connect cell centre $k$ to the points on the interface that are closest to cell centres $k$ and $l$, shown as $\vec{dv}_{kl,1}$ and $\vec{dv}_{kl,2}$ in Fig. \ref{vecdecomp}. $\vec v_{kl}$ is then constructed using a weighted sum of $\vec{dv}_{kl,1}$ and $\vec{dv}_{kl,2}$,
\begin{equation}
    \label{vkl}
    \vec v_{kl} = w_{k\sigma} \vec{dv}_{kl,1} + w_{l\sigma} \vec{dv}_{kl,2}.
\end{equation}

The vector $\vec v_{km}$ is constructed similarly but with vectors and weights associated with interface $\tau$ that connects cells $k$ and $m$. The interface normal is then decomposed into these two vectors by solving: 
\begin{equation}
    \label{normaldecomp}
    c_{kl} \vec v_{kl} + c_{km} \vec v_{km} = \vec n_\sigma 
\end{equation}
for the coefficients $c_{kl}$ and  $c_{km}$. These coefficients allow us to calculate the gradients across the interface without explicitly calculating the values of $u$ at the interpolation points. The one-sided flux $F_k$ is then calculated via the weighted sum of these gradients:
\begin{equation}
    \label{F_k}
    F_{k\sigma} = - D_k \left[ w_{l\sigma} c_{kl} (u_l - u_k) + w_{m\tau} c_{km} (u_m - u_k) \right].
\end{equation}

To calculate the net flux $\Tilde{F}_\sigma$ from Eqn. \ref{fluxes}, $F_{l\sigma}$ is constructed similarly to $F_{k\sigma}$ but from the perspective of cell $l$. For polar/height interfaces such as interface $\lambda$ in Fig. \ref{cellstencil}, the process is the same except that the stencil for constructing one-sided fluxes uses the interface in question and the anti-clockwise adjacent interface as opposed to clockwise.
\subsection{Gas}

In \texttt{cuDisc}, the 1D viscous evolution equation for gas surface density is solved, and the 2D gas distribution is calculated through hydrostatic equilibrium. Currently, there is no feedback from the dust to the gas; however, it may be added in the future. The viscous evolution equation is given by \citep{pringle1981}
\begin{equation}
    \label{1dgas}
    \frac{\partial \Sigma_g}{\partial t} = \frac{3}{R} \frac{\partial}{\partial R} \left(R^{1/2} \frac{\partial}{\partial R} (R^{1/2} \nu \Sigma_g)\right) + S_g,
\end{equation}
where $S_g$ represents any source terms (e.g. mass loss via photoevaporative winds). The equation is solved in two stages via operator splitting; in the first stage, the first term on the right-hand side is solved as a diffusion problem using an explicit forward-time-centered-spatial scheme, and in the second stage, the source terms are solved using explicit Eulerian updates. \\

The equations we use to calculate the gas velocities are found from viscous theory \cite[see e.g][]{urpin1984,balbus1999}, written in the coordinate grid basis of \texttt{cuDisc}, $(\nvec{r}, \nvec{Z})$. These velocities are needed explicitly for their influence on dust dynamics - they are not used for evolving the gas. Under the assumption that $v_Z \ll v_R, v_\phi$ (justified for thin discs as $v_Z \sim (H/R) v_R$ where $H$ is the gas scale height, see \citealt{urpin1984}), we find
\begin{equation}
    \label{vphigas}
    v_\phi = \bigg[\bigg.\frac{GM_* \cos^2{\theta}}{r} + \frac{r}{\rho_g}\left(\frac{\partial P}{\partial r} - \sin{\theta}\frac{\partial P}{\partial Z}\right)\bigg.\bigg]^{1/2},
\end{equation}

\begin{multline}
    \label{vRgas}
    v_R = r\cos{\theta} \left[\frac{\partial(rv_\phi)}{\partial r} - r\sin{\theta}\frac{\partial v_\phi}{\partial Z}\right]^{-1} \times\\
    \left\{\frac{1}{\rho_g} [\nabla \cdot \vec{T}]_\phi -v_{Z,\text{cyl}}\frac{\partial v_\phi}{\partial Z}\right\},
\end{multline}
where the derivatives are performed along the grid directions of spherical $r$ and cylindrical $Z$. $v_{Z,\text{cyl}}$ is the physical velocity in the $Z$-direction. \HL{We do not include a form for the vertical velocity due to viscosity here as it depends on assumptions made about the turbulence model \citep[see e.g.][]{philippov2017}; it can also be governed by other processes such as disc winds. For these reasons, by default (and for the tests in this paper) the vertical gas velocity is assumed to be 0. Different models of vertical gas motions can be included depending on the problem in question.} $\theta$ is the angle above the disc mid-plane, $P$ is the gas pressure, given by the ideal gas law, and $\vec{T}$ is the viscous stress tensor. To calculate $[\nabla \cdot \vec{T}]_\phi$, the $r\phi$ and $Z\phi$ components of $\vec{T}$ are required. Assuming a Navier-Stokes-like viscosity, these have the forms:
\begin{equation}
    \label{Trphi}
    T^{r\phi} = \frac{\rho_g \nu}{\cos^2{\theta}} \left[r\frac{\partial}{\partial r}\left(\frac{v_\phi}{r}\right) - \sin{\theta}\frac{\partial v_\phi}{\partial Z}\right],
\end{equation}
\begin{equation}
    \label{Tzphi}
    T^{Z\phi} = \frac{\rho_g \nu}{\cos^2{\theta}} \left[\frac{\partial v_\phi}{\partial Z} - r\sin{\theta}\frac{\partial}{\partial r}\left(\frac{v_\phi}{r}\right) \right],
\end{equation}
and the tensor divergence in the $\phi$ direction is explicitly written as
\begin{equation}
    \label{divT}
    [\nabla \cdot \vec{T}]_\phi = \frac{1}{r^3} \frac{\partial (r^3 T^{r\phi})}{\partial r} + \frac{\partial T^{Z\phi}}{\partial Z}.
\end{equation}

These derivatives are calculated using finite differencing on the grid variables to calculate cell-centred values for the gas velocities.

\subsection{Kinematic viscosity}

The kinematic viscosity $\nu$ is important for governing the gas' evolution and dust diffusion. It is often parameterised following \cite{shakira1973} as $\alpha c_s H$ where $H$ is the gas scale height, and $\alpha$ is a constant that is used to control the strength of the effective viscosity generated by turbulent processes. With a constant $\alpha$, this assumes that the viscosity varies with temperature changes in the disc. However, $\alpha$ need not be assumed constant; it can depend on other quantities such as ionisation fraction or magnetic field strength \cite[see e.g][]{jankovic2021}. A different way to parameterise the viscosity that removes this variation with the temperature profile calculated by \texttt{cuDisc} is $\nu= \nu_0 (R/R_0)$, where the linear dependence on radius comes from assuming a mid-plane temperature profile proportional to $R^{-1/2}$. Both methods of setting the viscosity are available in \texttt{cuDisc}. By default, $v_0$ is calculated assuming a mid-plane temperature of 100 K at 1 AU, but the user can edit this. 

\subsection{Hydrostatic equilibrium}

The 2D gas density structure is calculated by solving the hydrostatic equilibrium equation,
\begin{equation}
    \label{hydroeqbm}
    \frac{dP}{dZ} = - \frac{G M_*}{(R^2 + Z^2)^{\frac{3}{2}}} \rho_g Z.
\end{equation}

First, we rewrite the equation using the ideal gas law to obtain 
\begin{equation}
    \label{hydroeqbmrewrite}
    d \log P = \frac{G M_*}{c_s^2} d\left(\frac{1}{r}\right),
\end{equation}
where we have utilised $ZdZ/r^3 = dr/r^2 = - d(1/r)$, $r$ being the spherical radius, and $c_s$ is the isothermal sound speed. We then take the exponential of Eqn. \ref{hydroeqbmrewrite} and in each cell $i,j$ calculate
\begin{equation}
    \label{hedisc}
    \frac{P_{i,j}}{P_{i,j-1}} = \exp\left[GM_* \left< \frac{1}{c_s^2}\right>_{i,j} \left( \frac{1}{r_{i,j}} - \frac{1}{r_{i,j-1}} \right) \right],
\end{equation}
where $\left< \frac{1}{c_s^2}\right>_{i,j}$ is the average reciprocal of the sound speed squared, 
\begin{equation}
    \label{avcs}
    \left< \frac{1}{c_s^2}\right>_{i,j} = \frac{1}{2}\left[ \frac{1}{c_{s,i,j}^2} + \frac{1}{c_{s,i,j-1}^2}\right].
\end{equation}
The pressure in each cell relative to the cell at $j=0$ is then found by multiplicatively scanning over $Z$,
\begin{equation}
    \label{heprod}
    \frac{P_{i,j}}{P_{i,j=0}} = \prod_{k=1}^j \frac{P_{i,k}}{P_{i,k-1}}. 
\end{equation}
This method enforces positivity of the calculated pressure. The pressure is then converted to density and normalised via the gas surface density,
\begin{equation}
    \label{rhonorm}
    \rho_{g,i,j} = \frac{\Sigma_{g,i} (RdR)_i}{\sum_{k} \omega_{i,k} V_{i,k}} \omega_{i,j},
\end{equation}
where $\omega_{i,j}$ is the unnormalised density and $(RdR)_i$ is calculated as $(RdR)_i = \frac{1}{2} d(R^2)_i = \frac{1}{2} [(R^e_{i+1})^2 - (R^e_{i})^2]$.

\subsection{Dynamics Tests}

\begin{figure}
    \centering
    \includegraphics[width=0.99\columnwidth, trim={0.1cm 0 0 0},clip]{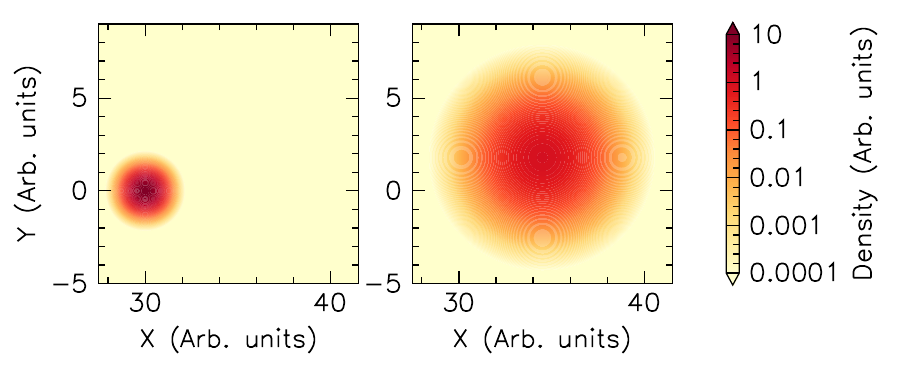}
    \caption{Advection-diffusion of a 2D gaussian pulse using the dynamics routine in \texttt{cuDisc}. The left and right panels show the initial and final states, respectively. This simulation had a resolution of 512 $\times$ 512 cells.}
    \label{2dgau}
\end{figure}

\begin{figure}
    \centering
    \includegraphics[width=0.95\columnwidth, trim={0.1cm 0 1.2cm 1cm},clip]{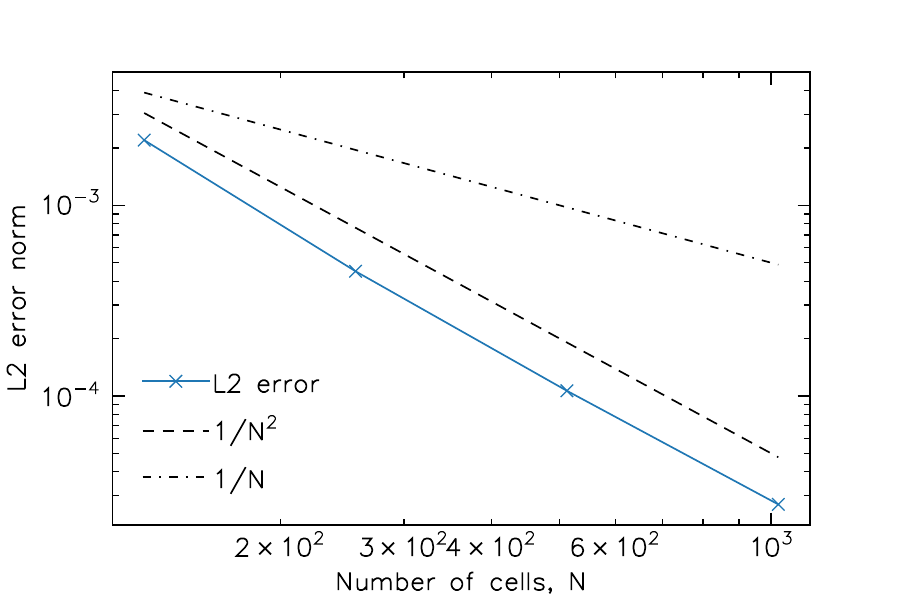}
    \caption{The L2 error norm for advection-diffusion of a 2D gaussian pulse using the dynamics routine in \texttt{cuDisc}. Slopes of $1/N$ and $1/N^2$, where the 2D grid has size $N\times N$, are shown for reference.}
    \label{l2error}
\end{figure}

To demonstrate that the code is second-order accurate, we now show test problems run using 2D Gaussian pulses with the analytic form
\begin{equation}
    \label{gaupulse}
    \rho(x,y,t) = \frac{A}{t} \exp \left(-\frac{(x-x_p(t))^2 + (y-y_p(t))^2 }{4Dt}\right),
\end{equation}
where $x_p(t) = x_0+v_x(t-t_0)$ with an equivalent expression for $y_p$. For these tests, we used the following parameter choices: $A=D=1$, $x_0=30$, $y_0=0$, $t_0=0.1$, $v_x=5$ and $v_y=2$. These pulses were advected in both $R$ and $Z$ directions whilst undergoing diffusion from $t=t_0$ to $t=1$. Cells were logarithmically-spaced in $R$ between 10 and 50 and linearly-spaced in $\theta$ between $-\pi/6$ and $\pi/6$. Cartesian areas and volumes were used, and the boundaries were set to outflow conditions. An example run can be seen in Fig. \ref{2dgau}. The L2 error norm, \HL{$\sqrt{\sum_{ij} (\rho_{ij} - \rho_{\text{an},ij})^2/(N\times N)}$,} was calculated using the analytic solution, $\rho_\text{an}$ compared to the computed solution, $\rho$. Fig. \ref{l2error} shows how this error decreases with number of cells $N$ on a 2D $N\times N$ grid. As shown, the L2 error is proportional to $1/N^2$, demonstrating that our method is second-order accurate. \\

\begin{figure}
    \centering
    \includegraphics[width=0.99\columnwidth]{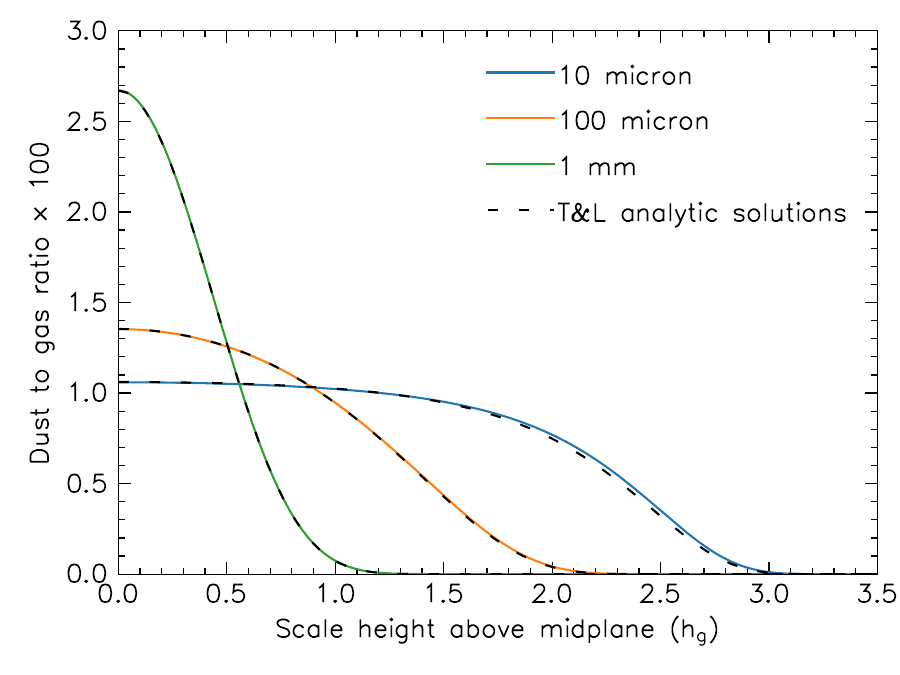}
    \caption{Steady-state dust-to-gas ratios as a function of height above the disc mid-plane (in units of the gas scale height, $h_g$) at a radius of 10 AU for three dust species. The analytic solutions given by \protect\cite{taklin02} are overplotted for comparison.}
    \label{taklin_dtg}
\end{figure}

\begin{figure}
    \centering
    \includegraphics[width=0.99\columnwidth]{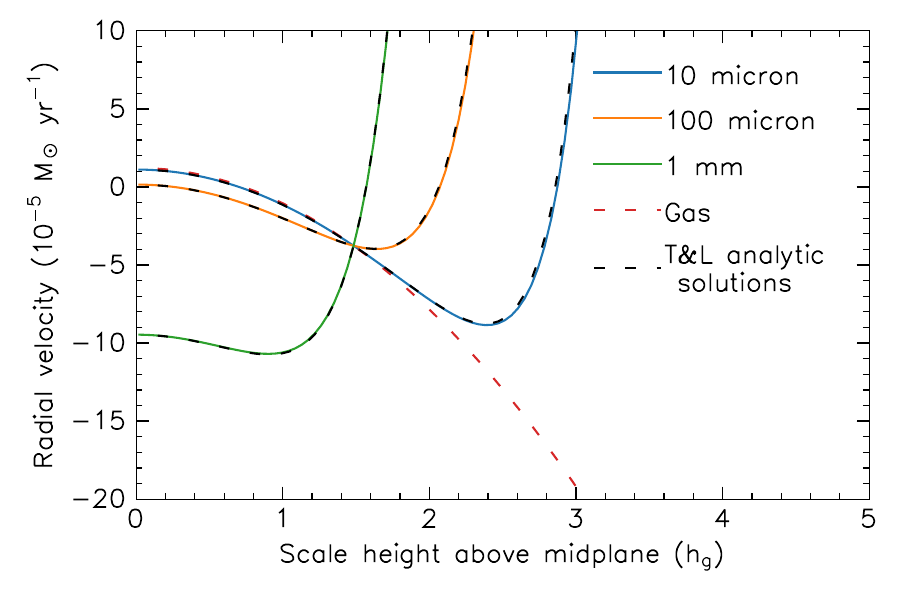}
    \caption{Steady-state radial velocities as a function of height above the disc mid-plane (in units of the gas scale height, $h_g$) at a radius of 10 AU for three dust species. The analytic solutions given by \protect\cite{taklin02} are overplotted for comparison.}
    \label{taklinrad}
\end{figure}

To test our code on disc-related problems, we compared simulated steady-state results to analytic results given by \cite{taklin02}. The gas was initialised using their analytic density and velocity profiles. Three dust species (10 $\mu$m, 100 $\mu$m and 1 mm) were initialised as being well-mixed with the gas (i.e. the same density and velocity profiles) with a standard dust-to-gas ratio of 0.01 applied to the density. In order to study vertical settling, radial fluxes were set to 0 for comparison with Takeuchi \& Lin's analytic profiles, and the code was run for 100,000 years to ensure a steady-state in the dust, \HL{as the settling timescale for the smallest grains in this set-up was 90,000 years}. \HL{300 cells were used in the vertical dimension.} Fig. \ref{taklin_dtg} shows the computed equilibrium dust-to-gas profiles of the dust species compared to the analytic solutions, showing strong agreement. \HL{A small discrepancy is visible for the smallest grains at large $Z$; this is due to our inclusion of the advection of momentum and because we do not make a thin disc approximation in the gravitational potential (Takeuchi \& Lin invoke $Z << R$ to simplify the problem).} With radial velocities turned on, we also match the analytic profiles for radial velocity as a function of height, as shown in Fig. \ref{taklinrad}.
\section{Temperature solver}\label{tempsolv}

Radiative transfer is a complex problem, and various numerical methods can be employed whose results differ depending on the weights attributed to accuracy versus computation time. For certain problems, calculated temperatures may need to be highly accurate and need not evolve (e.g. comparing line fluxes to observations), allowing long computation times to be a viable option. Other problems may not require such high accuracy in the calculation of temperature but instead may require fast calculation to assess the thermal evolution of a system. \\

Since the aim of \texttt{cuDisc} is to study the evolution of protoplanetary discs over long timescales ($\sim 10^5$ yr or longer), we require a fast temperature solver. For this reason, our method uses a hybrid of ray-tracing and multi-band flux-limited diffusion as opposed to exact techniques such as Monte Carlo. We benchmark our solver against \texttt{RADMC-3D} \citep{2012ascl.soft02015D}, a Monte Carlo code that requires longer computation times for more accurate results. 

\subsection{Hybrid ray-tracing + multi-band flux-limited diffusion}

In \texttt{cuDisc}, we use a hybrid radiative transfer method that employs ray-tracing from the star for heating due to stellar radiation and multi-band flux-limited diffusion (FLD) for treating re-emitted/scattered radiation. FLD is a model developed by \cite{1981ApJ...248..321L} that aims to solve the radiative transfer problem by treating the transport of radiative energy as a diffusion problem. The diffusive flux is limited to ensure that radiation does not propagate faster than the speed of light. The limiter is chosen to tend towards the correct physical forms in limits of high and low optical depth. The following scheme used in \texttt{cuDisc} is based on similar schemes detailed by \cite{kuiper2010, Commercon2011, bitsch2013}. In conservative form, the equations governing the internal energy density, $\epsilon$, and radiative energy density in a given wavelength band $b$, $E^b_{R}$, are given by
\begin{equation}
\label{inten}
    \frac{\partial\epsilon}{\partial t} = - \sum_b \left(\sum_s \rho_s \kappa^b_{P,s}\right) \left(f^b(T) B(T) - cE_R^b\right) + S_\text{heat},
\end{equation}
\begin{equation}
\label{raden}
    \frac{\partial E_R^b}{\partial t} + \nabla \cdot \vec{F}^b = \left(\sum_s \rho_s \kappa_{P,s}^b\right) \left(f^b(T) B(T) - cE_R^b\right) + S_{\text{sca}}^b,
\end{equation}
where $\rho_s$ is the mass density of species $s$, $\kappa_{P,s}^b$ is the Planck mean opacity in wavelength band $b$ for species $s$, $B(T) = 4\sigma T^4$ accounts for radiative cooling, $S_\text{heat}$ accounts for heating via stellar irradiation and viscous processes and $S_{\text{sca}}^b$ is the scattered stellar radiation. As Eqn. \ref{raden} is for a particular wavelength band, a Planck factor $f^b$ is included that accounts for the fraction of total radiated thermal energy that is emitted in the particular band in question. We do not include advective terms in our formulation. $\vec{F}^b$ is the flux of radiative energy in a given band, and this term is calculated in flux-limited diffusion as 
\begin{equation}
    \label{radflux}
    \vec{F}^b = - \frac{c\lambda^b}{\rho \kappa_R^b} \nabla E_R^b,
\end{equation}
where $\kappa_{R}^b$ is the Rosseland mean opacity in the given band and $\lambda^b$ the flux-limiter. In our multi-band scheme, the Planck opacities in Eqns. \ref{inten} \& \ref{raden} are calculated using absorption opacities, whilst the Rosseland opacity is calculated using the total opacity, absorption plus scattering. The choice of flux-limiter used for \texttt{cuDisc} is that given by \cite{1989A&A...208...98K}, 
\begin{equation}
    \lambda^b = 
    \begin{cases}
        \dfrac{2}{3 + \sqrt{9 + 10R_b^2}}, & \text{for } R_b < 2 \vspace{0.2cm}\\ 
        \dfrac{10}{9 + 10R_b + \sqrt{81 + 180R_b}}, & \text{for } R_b > 2
    \end{cases}
\end{equation}
where 
\begin{equation}
    R_b = \frac{1}{\rho \kappa_{R}^b} \frac{|\nabla E_{R}^b|}{E^b_{R}}.
\end{equation}
Taking the limits of high and low optical depth ($\rho\kappa_R \gg 1$ \& $\rho\kappa_R \ll 1$) the flux tends to the diffusion limit, ${\vec{F} = -(c/3\rho\kappa_R) \nabla E_R}$, and the free-streaming limit, ${\vec{F} = - c E_R (\nabla E_R / |\nabla E_R|)}$, respectively. \\

The heating term $S_\text{heat}$ is split into stellar heating and viscous heating. The viscous heating is calculated explicitly as 
\begin{equation}
    \label{vischeat}
    S_\text{visc} = \frac{9}{4} \alpha \rho_g c_s^2 \Omega,
\end{equation}
where $\alpha$ parameterises the effective turbulent viscosity of the disc \citep{shakira1973}, $c_s$ is the isothermal sound speed, and $\Omega$ is the Keplerian angular velocity. Stellar heating in each cell is calculated via ray tracing from the star using
\begin{equation}
    \label{stellheat}
    S_* = \sum_B \frac{L_B}{4 \pi r^2 \Delta r} e^{-\tau_B} (1-e^{-\rho \kappa_B \Delta r}) (1-a_B),
\end{equation}
where the sum is over the wavelength bands used for stellar radiation, $L_B$ is the stellar luminosity in each band, and $\Delta r$ is the (spherical) radial length of the cell. By default, the number of bands used for stellar heating ($\sim 100)$ is larger than the number used for the FLD routine detailed above ($\sim 20$) as the heating calculation is computationally cheap. For each wavelength band, the $e^{-\tau_B}$ term accounts for the flux that has been attenuated (removed through extinction; i.e. absorption plus scattering opacity) by the material along the radial path between the star and the cell of interest, whilst $(1-e^{-\rho \kappa_B \Delta r})$ accounts for the amount of remaining flux that is attenuated in the cell. $\kappa_B$ is the extinction of stellar radiation at the centre of each wavelength band. The term $(1-a_B)$ includes the albedo, $a_B$, to take into account the effect of scattering. The albedo is calculated as the ratio of scattering opacity to total extinction at each wavelength. The diffusive flux generated by scattering at each wavelength is therefore given by 
\begin{equation}
    \label{stellsca}
    S_{\text{sca},B} = \frac{L_B}{4 \pi r^2 \Delta r} e^{-\tau_B} (1-e^{-\rho \kappa_B \Delta r}) a_B.
\end{equation}
This is then binned into the wavelength bins used for the FLD routine and included in Eqn. \ref{raden}. \\

\begin{figure*}
    \centering
    \includegraphics[width = 0.98\textwidth]{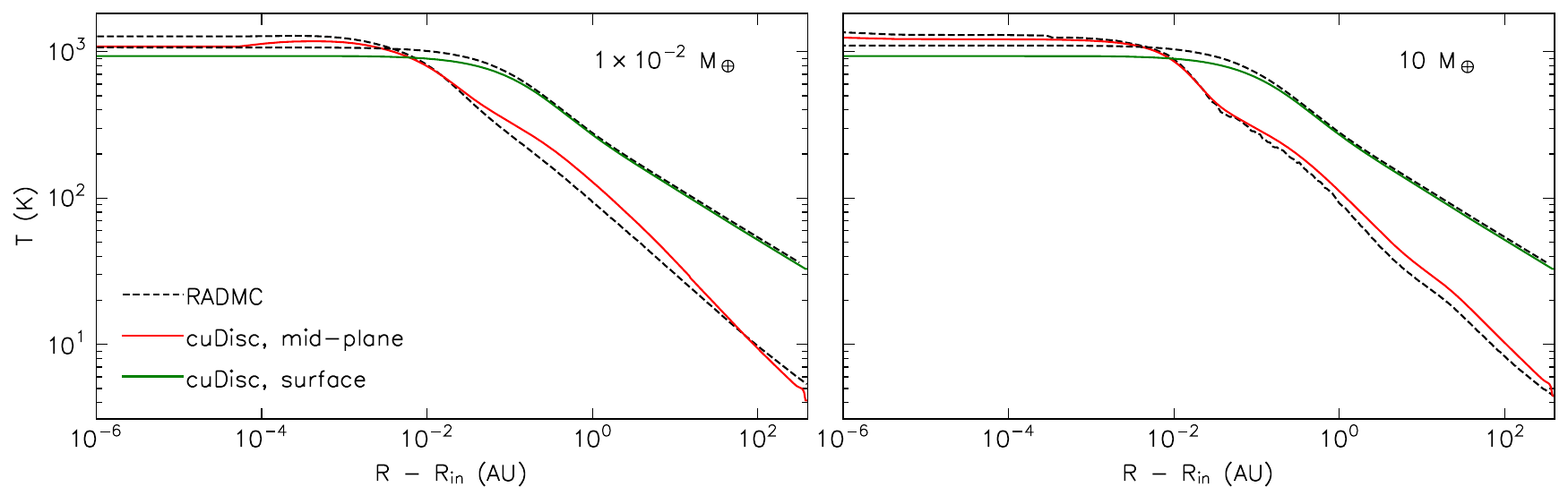}
    \caption{Comparison of mid-plane and surface temperatures calculated by \texttt{cuDisc} and \texttt{RADMC-3D} for the \protect\cite{pinte2009} benchmark adapted to have a grain distribution. The $x$-axis is plotted with units of cylindrical radius minus the inner disc radius (0.1 AU). The left and right plots show models with total dust masses of $1\times10^{-2}$ $M_\oplus$ and 10 $M_\oplus$, respectively.}
    \label{fldradmccomp}
\end{figure*}

The internal energy is given by ${\epsilon = \rho c_V T}$, where $\rho$ is the total density of all species and $c_V$ is the specific heat at constant volume, meaning that Eqn. \ref{inten} can be written in terms of $B$,
\begin{equation}
\label{intenB}
    \frac{\rho c_V}{16\sigma T^3}\frac{\partial B(T)}{\partial t} = - \rho \kappa_P (B(T) - cE_R) + S.
\end{equation}
This allows the system of equations to be set up and solved implicitly for the quantities $B$ and $E_R$. This method ensures the correct equilibrium is reached for large $\Delta t$ - this being the standard case for our simulations. We also note that the \cite{Commercon2011} linearization writes $B(T)$ in terms of $T$, instead of $T$ in terms of $B(T)$ as we do. The discretised forms of Eqns. \ref{intenB} \& \ref{raden} for cell $i,j$ at time-step $n$ are
\begin{multline}
    \label{flddiscint}
    \frac{\rho^n_{i,j} c_V}{16\sigma (T^n_{i,j})^3 \Delta t} B^{n+1}_{i,j}V_{i,j} \\
    \hspace{1cm}+ \sum_b \left(\sum_s \rho^n_{s,i,j} \kappa_{P,s,i,j}^b \right) \left(f^b_{i,j} B^{n+1}_{i,j} - cE^{b,n+1}_{R,i,j}\right) V_{i,j} \\
    = \frac{\rho^n_{i,j} c_V }{16\sigma (T^n_{i,j})^3 \Delta t} B^{n}_{i,j} V_{i,j} + S^n_{\text{heat},i,j} V_{i,j},  
\end{multline}
and 
\begin{multline}
    \label{flddiscrad}
    \frac{E^{b,n+1}_{R,i,j}}{\Delta t} V_{i,j}  -  \left(\sum_s \rho^n_{s,i,j} \kappa_{P,s,i,j}\right) \left(f^b_{i,j} B^{n+1}_{i,j} - cE^{b,n+1}_{R,i,j}\right) V_{i,j} + \\
    \sum^{i+1}_{k=i-1} \sum^{j+1}_{l=j-1} \tilde{D}^b_{kl} E^{b,n+1}_{R,k,l} = \frac{E^{b,n}_{R,i,j}}{\Delta t} V_{i,j} + S^{b,n}_{\text{sca},i,j} V_{i,j}.
\end{multline}
$\tilde{D}^b_{kl}$ is the diffusion matrix, constructed by summing all terms that apply to cell $k,l$ when finding $\Tilde{F}_\sigma A_\sigma$ (Eqn. \ref{fluxes}) for each of the four interfaces around cell $i,j$. The construction of the elements of $\tilde{D}_{kl}$ can be found in more detail in Appendix A. We time-lag the diffusion constant in Eqn. \ref{radflux} and the Planck factor, $f^b$ (i.e. they are evaluated at $t^n$ instead of $t^{n+1}$). The system of equations defined by Eqns. \ref{flddiscint} \& \ref{flddiscrad} are solved using incomplete-LU preconditioning and the Bi-Conjugate Gradient Stabilized (BiCGStab) solver described in \cite{nvidiabicstab}. The convergence criterion is set by comparing the maximum fractional residual between the left and right-hand sides of each equation at each iteration to a user-controlled relative tolerance, set by default to $10^{-4}$.

\subsection{Opacities}

Dust absorption and scattering opacities can be set in various ways in \texttt{cuDisc}. Simple, functional forms that approximate the opacity dependence on grain size and wavelength may be used, or opacity tables generated using packages such as the DSHARP opacity library \citep{DSHARP2018} can be read in. If tables are used, \texttt{cuDisc} can interpolate the opacity data to the user-defined grain sizes and wavelength bins using piecewise-cubic
Hermite interpolation \citep{fritsch80} in both grain size and wavelength space. In general, more wavelength bins are used for the cheaper stellar heating calculation in Eqn. \ref{stellheat} than are used in the FLD scheme. For conversion between these two wavelength grids, the opacities are binned by calculating the Planck mean opacity over the range of wavelengths corresponding to each coarse bin - these partial means result in a better estimate of the temperature in optically thin regions when a small number of bins are used. 

\subsection{Radiative transfer tests}

To test our temperature solver, we ran modified versions of standard benchmark tests used in the literature for comparing temperature solvers: the \cite{pascucci2004} and \cite{pinte2009} tests. Fig. \ref{fldradmccomp} shows the comparison of mid-plane and surface temperatures calculated using \texttt{cuDisc} and the Monte Carlo radiative transfer code \texttt{RADMC-3D} for the Pinte test, modified to include a distribution of grain sizes. The density of each grain species $i$ was set by

\begin{equation}
    \label{pintedensity}
    \rho_i(R,Z) = {F_i}\frac{\Sigma_{\text{tot}}(R)}{\sqrt{2\pi} h_i(R)} \exp\left(-\frac{Z^2}{2h_i^2(R)}\right), 
\end{equation}
where $F_i$ is the fraction of the total surface density $\Sigma_\text{tot}$ distributed to each grain size and $h_i$ the scale height for each grain. The grain size distribution was set according to the MRN \citep{MRN77} distribution with a maximum grain size of 0.5 cm, and the total surface density was set according to \cite{pinte2009} as 
\begin{equation}
    \label{pintesigma}
    \Sigma_\text{tot} = \Sigma_0 (R/\text{AU})^{-1.625},
\end{equation} 
where $\Sigma_0$ was set by the total desired disc mass. The scale height for each grain size was set as the approximate height reached once grains are settled \citep{dubrulle1995},
\begin{equation}
    \label{pintescaleheights}
    h_i = h_g \sqrt{\frac{1}{1+\text{St}_i/\alpha}},
\end{equation}
where $h_g$ is the gas scale height, St$_i$ is the grain Stokes number, set simply in this problem as St$_i = 0.05 a_i (R/\text{AU})$ where $a_i$ is the grain radius, and $\alpha$ is the gas turbulence parameter, set to $1\times10^{-3}$. The gas scale height was set according to \cite{pinte2009} as $h_g = 10 (R/100 \text{ AU})^{1.125}$ AU. The opacities of the dust grains were set using the \cite{DSHARP2018} tool for calculating opacities, together with the dielectric constants given by \cite{Draine_2003}. Scattering was treated as being isotropic as \texttt{cuDisc} is currently unable to treat anisotropic scattering. Fig. \ref{fldradmccomp} shows the results for two different total disc masses ($1\times10^{-2}$ \& 10 $M_\oplus$). These masses corresponded to mid-plane optical depths from the star to the observer at 0.81 $\mu$m of 87 and $8.7\times10^4$, respectively. These values are just over an order of magnitude smaller than the optical depths for the Pinte test at the same disc masses, as our use of a grain distribution lowers the overall opacity of the disc due to the presence of large grains. The inner and outer bounds of the grids were set to (0.1, 400) AU in $R$ and (0, $\pi/4$) rad in $\theta$. 150 equally-spaced cells were used in $\theta$, whilst the number of cells in $R$ was adapted depending on the disc mass. For both disc masses, 200 logarithmically-spaced cells were used between 0.15 \& 400 AU, whilst to minimise the optical depth in cells very close to the inner edge, we follow the method detailed by \cite{ramsey2015} and use 40 \& 96 logarithmically-spaced cells between 0.1 \& 0.15 AU for the lower \& higher disc mass respectively. 100 logarithmically-spaced wavelengths between 0.1 \& 3000 $\mu$m  were used for stellar heating and binned down to 20 bands for the FLD calculations. Apart from the binned wavelength bands (which are not required), these same grid structures were used in \texttt{RADMC-3D} for consistency. Resolution tests were performed, which confirmed that convergence was reached for the grid resolutions quoted here. \\ 

We find that our results in the optically thin surface layers are accurate to within a few percent for the entirety of the disc apart from the very inner region between 0.1 and 0.2 AU. At the mid-plane, the accuracy varies but remains within $\sim$20 \% for most of the disc - a level of accuracy high enough for our problems.

\section{Dust growth and fragmentation}

Dust coagulation is performed using the method outlined in \cite{brauer2008} but without the vertical integration that they use to convert the problem to 1D along the disc mid-plane. In this method, the Smoluchowski coagulation equation \citep{smol1916} becomes 
\begin{equation}
    \dot{\rho}_k = \sum_{i=0}^{N-1} \sum_{j=0}^i \dot{\rho}_{\text{gain}, ijk} - \sum_{i=0}^{N-1} \dot{\rho}_{\text{loss}, ik}
    \label{smol}
\end{equation}
where dots represent time derivatives, $N$ is the total number of dust species, $\rho_k$ is the density of the $k$-th dust species, and $\dot{\rho}_{\text{gain}, ijk}$ and $\dot{\rho}_{\text{loss}, ik}$ are the gains in density due to collision products of other grains (through coagulation and fragmentation) and losses in density due to collisions between the $k$-th dust species and all others respectively. The loss is given by
\begin{equation}
    \dot{\rho}_{\text{loss}, ik} = m_k n_i n_k K_{ik} (p_{{\rm coag},ij} + p_{{\rm frag},ij})
\end{equation}
where $m_k$ is the mass of the $k$-th grain, $K_{ij}$ is referred to as the kernel, and $p_{{\rm coag},ij}$ and $p_{{\rm frag},ij}$ are the probabilities of coagulation and fragmentation, which follow 
\begin{equation}
    p_{\text{coag},ij} + p_{\text{frag},ij} = 1
\end{equation}
in the absence of bouncing (as assumed here). The form of this kernel is taken from \cite{birnstiel2010} and is given by
\begin{equation}
    \label{kernel}
    K_{ij} = \sigma_{ij} \Delta v_{ij},
\end{equation}
where $\sigma_{ij}$ is the collision cross-section of the grains and $\Delta v_{ij}$ is the relative velocity of the grains, given by
\begin{equation}
    \Delta v_{ij}^2 = v_{\text{BM}, ij}^2 + v_{\text{turb}, ij}^2 + (\Delta v_{\text{lam}, ij})^2, 
\end{equation}
where $v_{\text{BM}, ij}$, $v_{\text{turb}, ij}$, and $\Delta v_{\text{lam}, ij}$ are the relative velocities due to Brownian motion, gas turbulence, and laminar dust motion respectively. Here $\Delta v_{\text{lam}, ij}$ is the root-mean-square relative velocity between the dust particles, calculated using the velocities determined by the dynamical solver. Brownian motion most strongly affects small grains and is given by
\begin{equation}
    v_{\text{BM}, ij}= \sqrt{\frac{8k_B T (m_i+m_j)}{\pi m_i m_j}},
    \label{vbrown}
\end{equation}
where $k_B$ is the Boltzmann constant and $T$ the disc temperature. The form of $v_{\text{turb}, ij}$ is taken from \cite{ormel2007} and is given by 
\begin{equation}
    v_{\text{turb}, ij} = \sqrt{\alpha} c_s V_{\text{rel}, ij},
\end{equation}
where $V_{\text{rel}, ij}$ is the relative velocity between grains scaled to the eddy velocity $\sqrt{\alpha} c_s$, 
\begin{equation}
    \label{OCVrel}
    V_{\text{rel}, ij} = \sqrt{\Delta V_\text{I}^2 + \Delta V_\text{II}^2},
\end{equation} 
where $\Delta V_\text{I}^2$ \& $\Delta V_\text{II}^2$ are the relative velocities induced by ``slow'' class I and ``fast'' class II eddies, respectively given by Eqns. 17 \& 18 in \cite{ormel2007}. For computational speed, we use a quadratic fit to find the Stokes number of the boundary between class I \& II eddies for a given pair of grains, St$^*$ (Eqn. 21d in \cite{ormel2007}). This fit is accurate to within 2\%. \\

In a collision, grains can coagulate to form more massive grains or fragment to produce a remnant grain and a set of fragments that are distributed over lower-mass grains. Combining all of these collision products gives us the density gain for a dust species as
\begin{multline}
    \dot{\rho}_{\text{gain}, ijk} = n_i n_j K_{ij} \left[p_{\text{coag},ij} (m_i + m_j) C_{ijk} \right. \\
    \left. + p_{\text{frag},ij} (m_{\text{rem},ij}R_{ijk} + m_{\text{frag},ij}F_{ijk})\right] ,
    \label{ngain}
\end{multline}
where $C_{ijk}$, $R_{ijk}$ and $F_{ijk}$ are the fractions of the mass associated with coagulation products, remnant products and fragment products that are distributed into bin $k$, and $m_{\text{rem},ij}$ and $m_{\text{frag},ij}$ are the masses of remnants and fragments produced by the collision. Note to avoid double counting we replace $K_{ij} \rightarrow \frac{1}{2}K_{ij}$ when $i=j$.\\

The particle velocities are assumed to follow a Maxwell-Boltzmann distribution around $\Delta v_{ij}$ to calculate the fragmentation probability. This is a simplified version of the calculations done by \cite{garaud2013}, who separated the stochastic (turbulent and Brownian) motions from the laminar motions; we are therefore assuming that stochastic motions are larger than any laminar relative motions. This is also the approach taken by \cite{Stammler_2022}. In this way, the fragmentation probability is given by
\begin{equation}
    p_{\text{frag},ij} = \left[\frac{3}{2}\left(\frac{v_\text{frag}}{\Delta v_{ij}}\right)^2+1\right] \exp\left[-\frac{3}{2}\left(\frac{v_\text{frag}}{\Delta v_{ij}}\right)^2\right].
    \label{p_frag}
\end{equation}
The coefficients $C_{ijk}$, $R_{ijk}$ and $F_{ijk}$ add the collision products into the appropriate mass bins. Since our standard choice of the mass grid is logarithmic, these products do not map directly to other mass bins and must be split over the two bins on either side of the product mass. Defining the integers $\ell(m)$ and $u(m)$ as: $m_{\ell(m)}$ being the largest mass such that $m_{\ell(m)} < m$ and $m_{u(m)}$ being the smallest mass such that $m_{u(m)} > m$, we may write 
\begin{equation}
    C_{ijk} = 
    \begin{cases}
        \varepsilon(m_i + m_j) & k = \ell(m_i + m_j)\\
        1-\varepsilon(m_i + m_j) & k = u(m_i + m_j) \\
        0 & \text{otherwise,} 
    \end{cases}
    \label{Cijk}
\end{equation}
and
\begin{equation}
    R_{ijk} = 
    \begin{cases}
        \varepsilon(m_{\text{rem},ij}) & k = \ell(m_{\text{rem},ij})\\
        1-\varepsilon(m_{\text{rem},ij}) & k = u(m_{\text{rem},ij}) \\
        0 & \text{otherwise,} 
    \end{cases}
    \label{Rijk}
\end{equation}
where 
\begin{equation}
    \varepsilon(m) = \frac{m_{u(m)} - m}{m_{u(m)}-m_{\ell(m)}}.
\end{equation}
The distribution of fragments, $F_{ijk}$, is calculated following \cite{rafikov20}. First, we assume that the number density distribution of fragments is given by a power law, 
\begin{equation}
    n(m)dm \propto m^{-\eta} dm
    \label{fragdist}
\end{equation}
where $n(m)dm$ is the number of particles per unit volume in the mass range $[m, m+dm]$ and $\eta$ is the fragmentation parameter that controls how fragments are distributed amongst the smaller mass grains; by default, we take the standard value of 11/6 found in the steady-state solutions of \cite{1969JGR....74.2531D,TANAKA1996450}. Converting to mass density and integrating over the bins to find the fraction of mass distributed into each bin gives
\begin{equation}
    \label{Nijk}
    N_k(m_{\rm max}) = \frac{\left(m^e_{k+1}\right)^{2-\eta} - \left(m^e_{k}\right)^{2-\eta} }{m_{\rm max}^{2-\eta} - \left(m^e_0\right)^{2-\eta}},
\end{equation}
where $m_{\rm max}$ is the mass of the largest fragment and where $m^e_{k+1}$ and $m^e_{k}$ are the upper and lower edges of the $k^{\text{th}}$ mass bin. \\

We then place each $m_{\text{frag},ij}$ into the bin $l$ by finding the smallest $l$ such that $m^e_{l+1} > m_{\text{frag},ij}$ and setting $m_{{\rm max},ij} = m^e_{l+1}$. From this we arrive at $F_{ijk} = N_k(m_{{\rm max},ij})$. However, $F_{ijk}$ is not used directly. Instead, we first determine the total amount of fragments falling into each $l$-bin before distributing them according to $N_k(m^e_{l+1})$. This results in a much more efficient algorithm, as outlined in \citet{rafikov20}.\\

The proportion of mass ending up as fragments is defined as a multiple of the mass of the smaller of the two colliding grains. Denoting the more massive particle as the target with mass, $m_\text{tar}$ and the smaller as the impactor with mass, $m_\text{im}$, we define the $\chi_\text{im}$ such that $m_{\text{frag},ij} = \text{min}\left(\chi_\text{im} m_\text{im},m_\text{tar}\right)$. The remnant mass is then given by
\begin{equation}
    m_{\text{rem}} = 
    \begin{cases}
        m_\text{tar} - \chi_\text{im} m_\text{im} & \text{for } \chi_\text{im} m_\text{im} < m_\text{tar} \\
        0 & \text{otherwise},
    \end{cases}
    \label{fragrem}
\end{equation}
where $\chi_\text{im}$ is a factor controlling the amount of material the impactor can remove relative to its mass, usually taken to be unity. The fragment mass is then given by the total mass of the colliding grains minus the remnant mass. \\

To integrate Eqn. \ref{smol}, \texttt{cuDisc} employs the Bogacki-Shampine 3(2) embedded Runge-Kutta method with error estimation \citep{BOGACKI1989321} that adaptively calculates the time-step required to meet user-specified relative and absolute tolerances. As default, these are a 1\% relative tolerance and an absolute tolerance of $1\times 10^{-10}$ multiplied by the sum over all grain densities in a cell. 

\subsection{Coagulation tests}

\begin{figure}
    \centering
    \includegraphics[width=0.99\columnwidth]{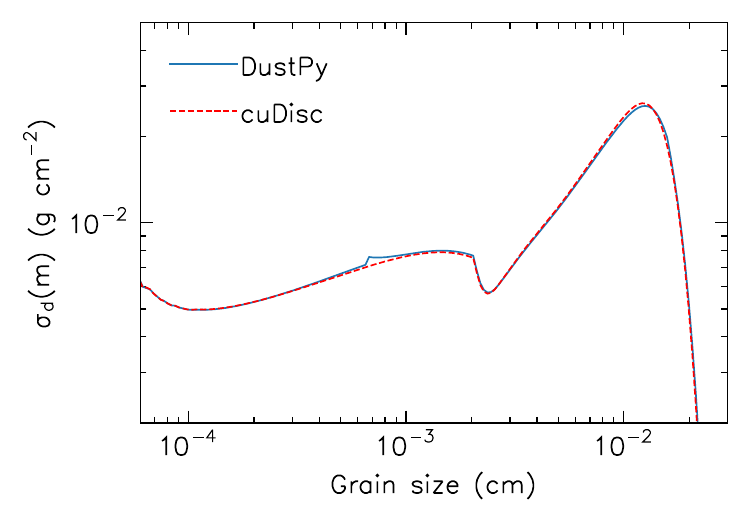}
    \caption{Comparison of steady-state dust grain size distributions calculated by \texttt{cuDisc} and \texttt{DustPy}. The simulation parameters used were $R = 20$ AU, $T = 36$ K, $\mu = 2.4$, $\alpha = 10^{-3}$ and $v_\text{frag} = 100$ cm s$^{-1}$.}
    \label{dustpycomp_0D}
\end{figure}

To test our coagulation routine, we produced direct comparisons to the 1D disc evolution code \texttt{DustPy} using a vertically integrated kernel \citep{birnstiel2010}, which can be seen in Fig. \ref{dustpycomp_0D}. The vertically integrated kernel is used to study the growth and fragmentation of dust surface densities instead of volume densities and is constructed by dividing the coagulation kernel (Eqn. \ref{kernel}) by $\sqrt{2\pi (h_i^2+h_j^2)}$, where $h_i$ \& $h_j$ are the scale heights of the dust grains, assuming a Gaussian vertical density profile. The relative vertical velocity is set according to the difference in terminal settling velocities of the dust grains in question, taking the dust grain scale height as its representative height $Z$ above the mid-plane. Each simulation was run with dynamics turned off, at a single radius in the disc, with the same initial conditions. The simulations were run until a steady-state was reached in the dust grain size distribution. To compare the two codes, we plot the mass-grid independent densities for each grain, given by
\begin{equation}
    \label{dustsigma1}
    \sigma_d(m) = m \frac{\partial \Sigma_d}{\partial m}.
\end{equation}
200 grain species were used for both simulations, as this is comparable to the typical amount used for science calculations. The two codes show strong agreement. The only slight difference at $\sim 7\times10^{-4}$ cm is due to different methods used by the two codes to fit the relative turbulent velocities described in \cite{ormel2007}. The difference occurs because the approximation used in \texttt{DustPy} is not continuous in this region. \\

We also ran tests using a constant kernel, the results of which were consistent with the analytic solutions.


\section{Code structure}

\begin{figure}
    \centering
    \includegraphics[width=0.92\columnwidth]{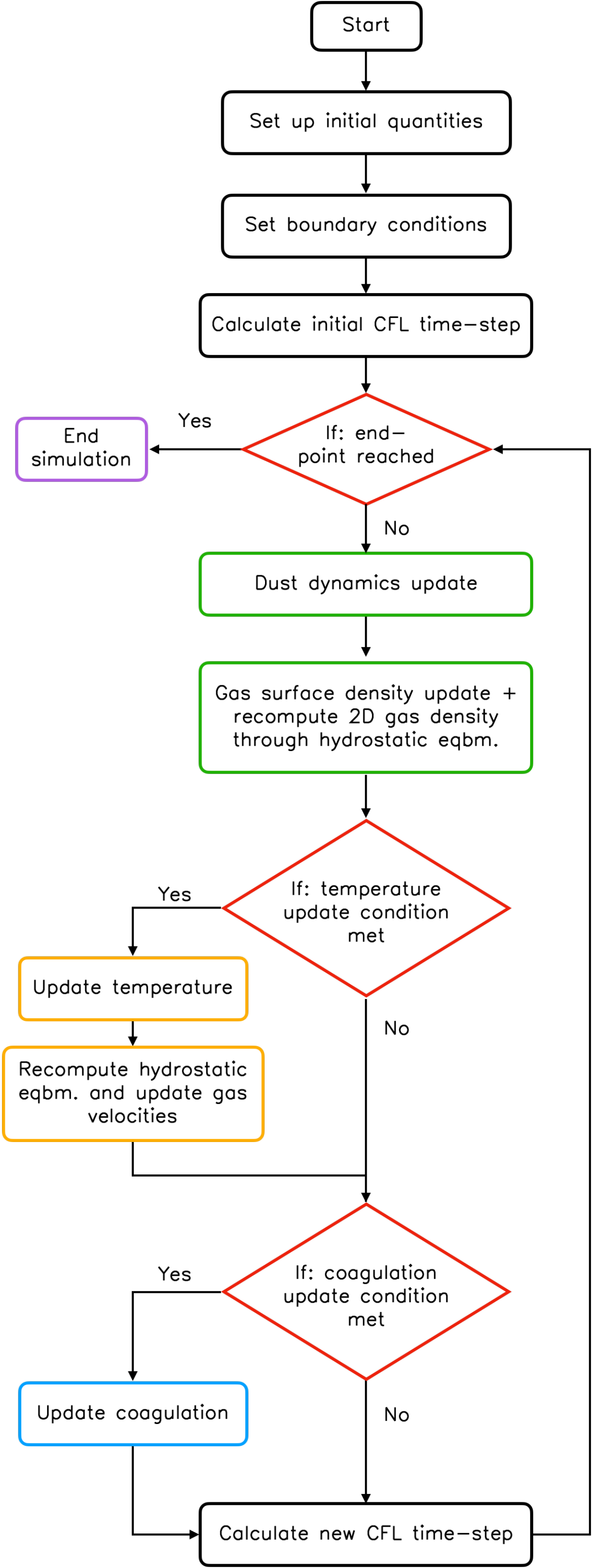}
    \caption{The schematic for \texttt{cuDisc}. Dust \& gas dynamics are updated every time-step whilst temperature and coagulation updates are performed when certain conditions are met. These conditions can be controlled by the user in order to align with the timescales relevant to the physical problem at hand.}
    \label{schem}
\end{figure}

The overall time evolution is controlled by the CFL condition on the dust dynamics. Gas, temperature and coagulation updates are then performed when required depending on the evolution time-scales associated with each physical process. A schematic of the code structure can be seen in Fig. \ref{schem}. By default, gas surface density updates and subsequent hydrostatic equilibrium calculations are also performed at each CFL time-step due to their low computational cost. For temperature calculations, an update is performed on the first time-step of the simulation, and the percentage change with respect to the initial state is calculated. How often the temperature solver should be employed is then controlled by choosing a percentage tolerance for temperature updates that allows the code to calculate the approximate time period that can be allowed to elapse before the next update is required; by default, this is set to 0.1\%. However, this tolerance should be chosen depending on the expected time-scales of temperature evolution for each specific problem. Coagulation updates are managed differently, given that the coagulation routine is explicit and, as such, performs its own sub-steps within each global time-step (i.e. sub-cycling). To control how often these updates are performed, the final sub-time-step within the coagulation routine is extracted and used to compare to global time-steps. By default, the coagulation routine is triggered when one sub-time-step taken from the last coagulation step has elapsed in global simulation time; however, this can be changed by the user.
\section{Grain growth in a steady-state transition disc}{\label{steadydisc}}

To show a simple science test case and compare it to other works, we ran simulations of a disc with an inner hole evolving towards a steady-state, representative of a ``transition'' disc structure \citep[e.g.][]{Owen2016}. To compare to \texttt{DustPy}, one simulation was run with a vertically isothermal temperature profile set to the mid-plane temperature adopted by \texttt{DustPy}, whilst one had the 2D temperature solver switched on. We will refer to these simulations as isothermal and non-isothermal from here on. The gas surface density was set to have a mass within 100 AU of 0.017 $M_\odot$ with a power law radial dependence and a sharp exponential cut off at 5 AU; i.e. $\Sigma_g = \Sigma_0 R_\text{AU}^{-1} \exp{[-(5/R_{\text{AU}})^{10}]}$. For these tests, the gas surface density was not evolved; however, the vertical gas density profile was updated for the non-isothermal test to maintain hydrostatic equilibrium. Radial gas velocities were set to 0, whilst azimuthal velocities were updated with the temperature to maintain force balance. 150 logarithmically-spaced cells were used in $R$ between 3 \& 20 AU, and a total of 150 cells were used in $\theta$ between 0 \& $\pi/4$, with 92 linearly-spaced cells between 0 \&  $\pi/9$ with double the resolution of the remaining linearly-spaced 58 cells. This sub-division was used to enhance the resolution at the mid-plane. The viscosity, $\nu$, was set according to \cite{shakira1973} via $\alpha c_s H$. The parameters chosen for each of the simulations are given in Table~\ref{params}. The stellar radiation was assumed to be a blackbody at the effective temperature given in Table~\ref{params}. Note that simulations were run for fragmentation velocities of 100 and 1000~cm~s\minone{}. For the 100~cm~s\minone{} runs, 100 dust species with grain sizes logarithmically spaced between 0.1 $\mu$m and 0.5 cm were initialised with a size distribution according to the MRN distribution \citep{MRN77} and spatially distributed as being well-mixed with the gas at a local dust-to-gas ratio of 0.01. For the 1000~cm~s\minone{} runs, the mass grid was adjusted to 135 logarithmically spaced grain sizes between 0.1 $\mu$m and 20 cm to account for the larger grains present. \HL{100 \& 135 mass bins were used respectively to give $m_{n+1}/m_n = 1.38$ for both sets of simulations. This choice maintains $>7$ bins per mass decade, a requirement for accuracy in the coagulation routine \citep{ohtsuki1990}. For these calculations, changes in dust composition throughout the disc due to sublimation were neglected.} In this section we discuss the 100~cm~s\minone{} runs, whilst the 1000~cm~s\minone{} runs are discussed in Section \ref{particlesweeping}. \\

\begin{figure*}
    \centering
    \includegraphics[width=1\textwidth]{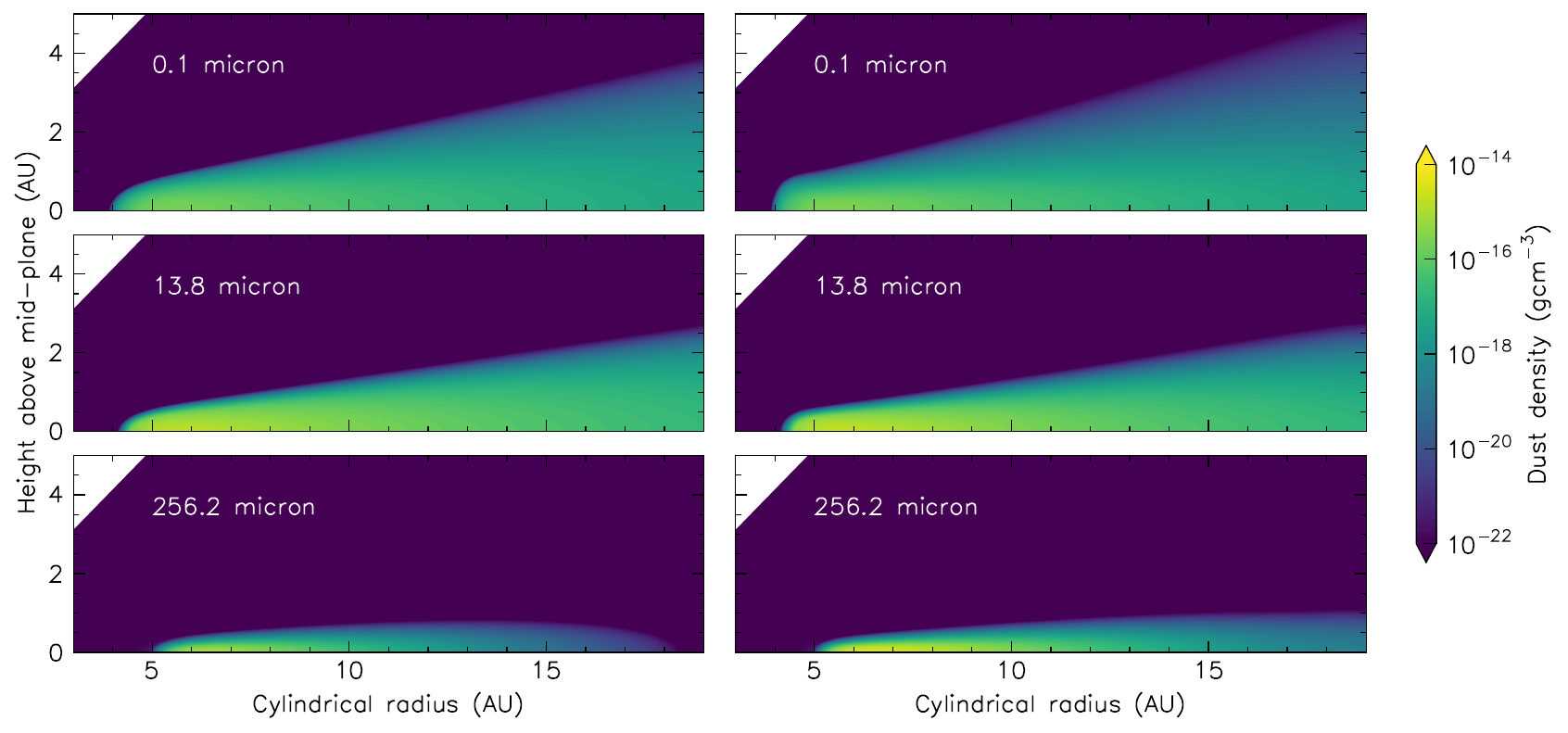}
    \caption{Dust profiles for three grain sizes after 1 Myr of evolution for a vertically isothermal disc (left) and a disc with the 2D temperature solver switched on (right). These runs were set with $v_\text{frag}=100$~cm~s\minone{}.}
    \label{dust2D}
\end{figure*}

\begin{figure}
    \centering
    \includegraphics[width=1\columnwidth]{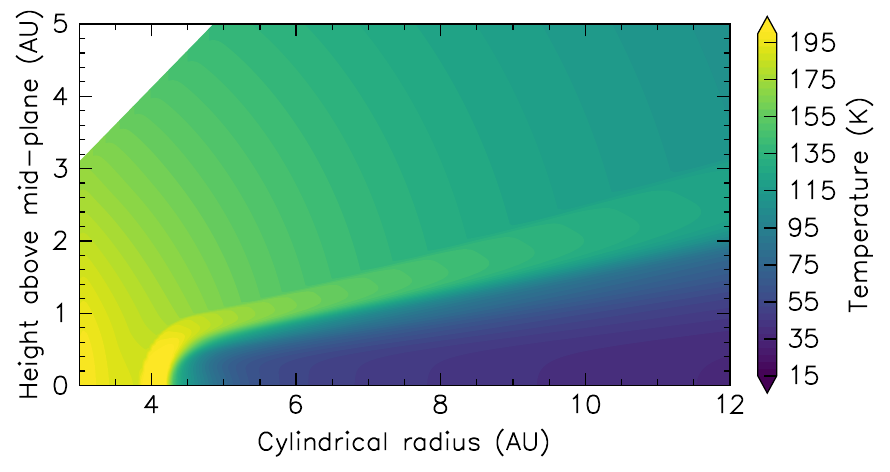}
    \caption{Temperature profile after 1 Myr of evolution for a disc with the 2D temperature solver switched on and $v_\text{frag}=100$~cm~s\minone{}. }
    \label{temp2D}
\end{figure}

Fig. \ref{dust2D} shows the 2D dust density profiles of three grain sizes after 1 Myr of evolution, at which time the simulations had reached a steady state, for both the isothermal disc and non-isothermal disc, whilst Fig. \ref{temp2D} shows the temperature profile of the non-isothermal disc after 1 Myr. Dust settling is apparent from the stratification of the different grains. The non-isothermal disc also exhibits an increase in the proportion of large grains due to the cool interior that allows for the coagulation of grains up to larger sizes. This is because the strength of the gas turbulent velocity is proportional to the sound speed when assuming a \cite{shakira1973} viscosity, and reduced turbulent velocities allow particles to grow larger before reaching the fragmentation limit. The temperature structure of the non-isothermal disc also exhibits a super-heated surface layer with temperatures greater than the blackbody equilibrium temperature; this is due to the small grains that occupy the upper regions of the disc whose opacities mean they are good absorbers of stellar optical photons, but bad emitters of their own infra-red photons \citep[see e.g.][]{chiang1997}. Above the super-heated layer, the dust densities drop to the floor value and the temperature is set by the opacities given to the gas. By default, these opacities are set as the dust-to-gas ratio floor value multiplied by the opacities of the smallest dust grains, as this leads to a smooth temperature transition in the upper regions of the disc. In reality, the temperature in the gaseous atmosphere above the dust disc is controlled by other processes \citep[see e.g.][]{woitke2009}. In principle, these processes can be included in the framework if necessary. \\

\begin{table}
    \centering
    \begin{tabular}{cc}
        \hline Parameter & Value  \\ \hline
        $\alpha$ & $10^{-3}$ \\
        $\mu$ & 2.4 \\
        $v_\text{frag}$ & 100, 1000~cm~s\minone{} \\
        $M_*$ & 1 $M_\odot$ \\
        $R_*$ & 1.7 $R_\odot$ \\
        $T_{*,\text{eff}}$ & 4500 K \\ 
        $M_\text{disc}$ & 0.017 $M_\odot$ \\ 
        \HL{$\Sigma_0$} & \HL{250 g~cm$^{-2}$} \\
        \HL{$\rho_m$} & \HL{1.6 g~cm$^{-3}$}\\
        Dust opacities & DSHARP mix \\ 
        & \citep{DSHARP2018} \\ \hline
        
    \end{tabular}
    \caption{Parameters used for the transition disc simulations detailed in Section \ref{steadydisc}.}
    \label{params}
\end{table}

\begin{figure}
    \centering
    \includegraphics[width=1\columnwidth]{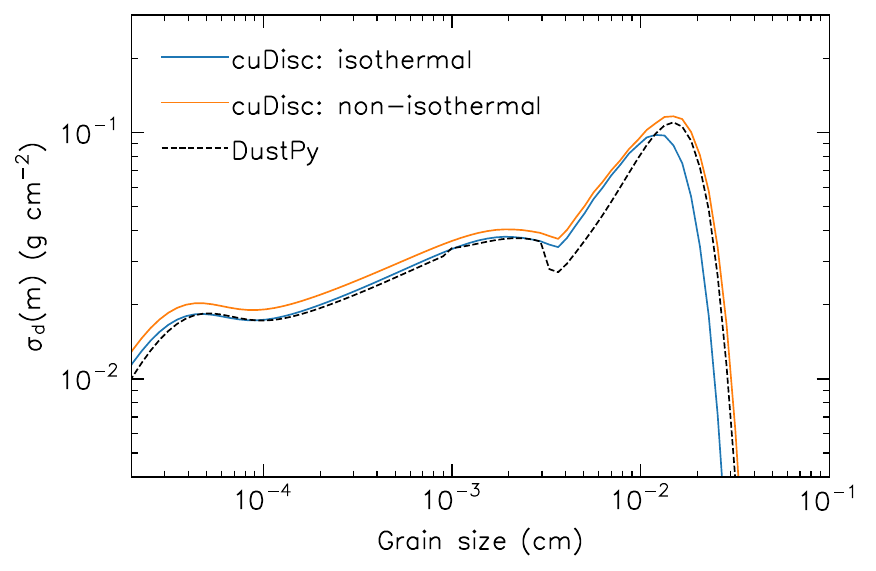}
    \caption{Vertically-integrated mass-grid independent dust densities at a radius of 6 AU for each $v_\text{frag}=100$~cm~s\minone{} simulation.}
    \label{graindist1D}
\end{figure} 

\begin{figure*}
    \centering
    \includegraphics[width=1\textwidth]{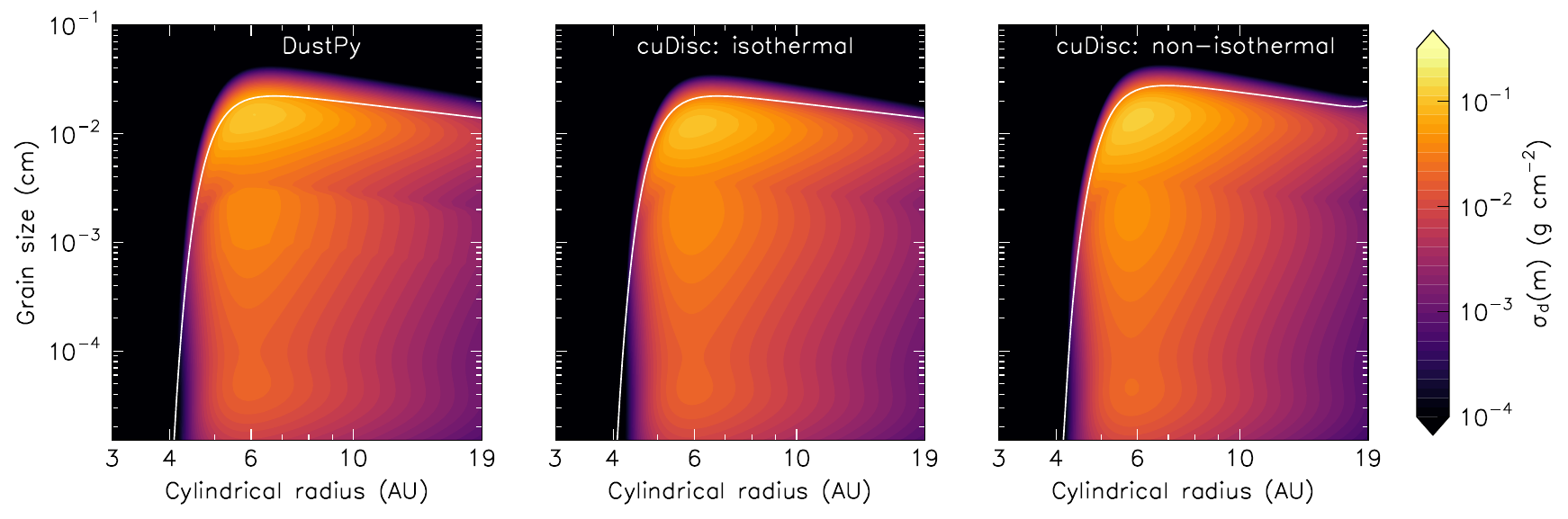}
    \caption{Dust grain size distributions as a function of disc radius for each $v_\text{frag}=100$~cm~s\minone{} simulation. The fragmentation limit (given by \citealt{birnstiel2012}) is over-plotted on each distribution for reference.}
    \label{graindist2D}
\end{figure*}

Figs. \ref{graindist1D} \& \ref{graindist2D} compare the vertically integrated dust grain size distributions for the different simulations at a radius of 6 AU and for the entire disc, respectively. Vertically-integrated mass-grid independent densities were calculated from the volume densities computed by \texttt{cuDisc} through
\begin{equation}
    \label{dustsigma}
    \sigma_d(m) = \int_{-\infty}^{\infty} m \frac{\partial \rho_d}{\partial m} dz.
\end{equation}
Fig. \ref{graindist2D} shows that the spatial distribution of dust is very similar for all runs; the maximum grain size is set by fragmentation throughout the entire disc due to the fairly low choice of 100 cm s$^{-1}$ as the fragmentation velocity. The non-isothermal run is slightly more peaked at 6 AU. This is because the equilibrium spatial distribution is set by the balance between diffusion and radial drift, and, as previously mentioned, the non-isothermal disc has lower turbulence and, therefore, less diffusion of the dust out of the dust trap. Fig. \ref{graindist1D} shows that the simulations follow very similar distributions up to $\sim30 \mu$m, with the non-isothermal run having an overall higher density across all grain sizes due to the increased surface density at the dust trap. The \texttt{DustPy} simulation exhibits a much more prominent trough at 30 - 40 $\mu$m than the \texttt{cuDisc} simulations, and the upper cut-off in the size distribution differ across the simulations; the \texttt{DustPy} and non-isothermal runs show similar cut-offs whilst the isothermal run is lower. 

\begin{figure*}
    \centering
    \includegraphics[width=1\textwidth]{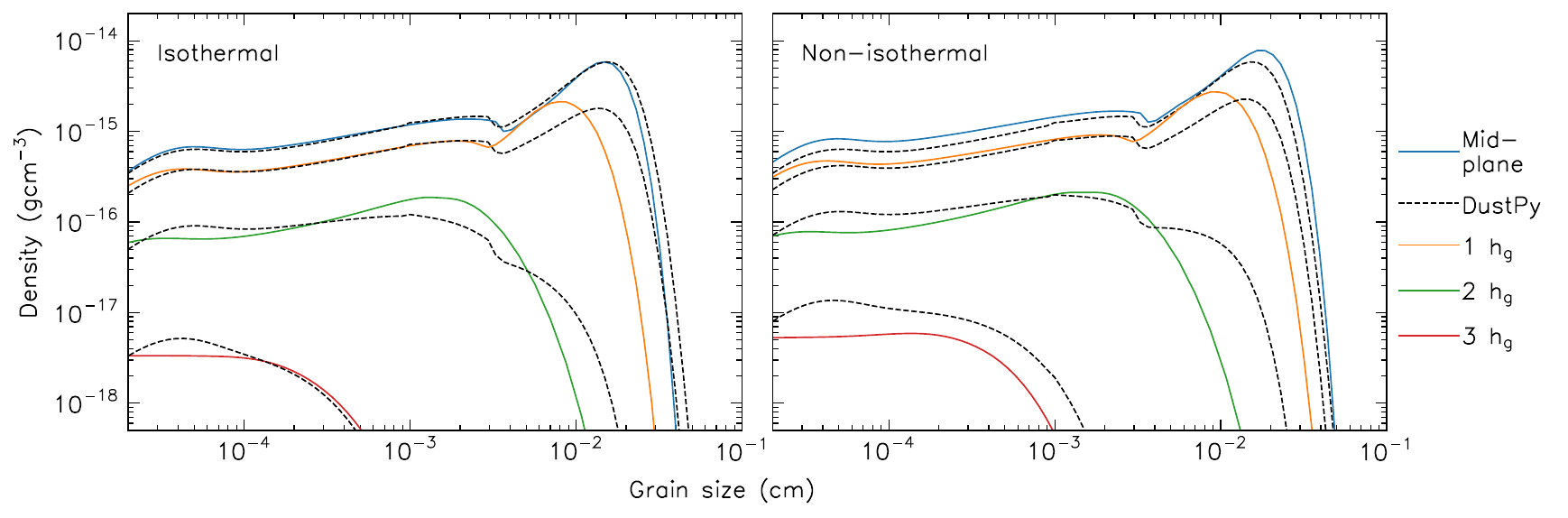}
    \caption{Dust grain size distributions at 1, 2 and 3 gas scale heights above the mid-plane for each $v_\text{frag}=100$~cm~s\minone{} simulation. The heights in $Z$ of the plotted gas scale heights in each plot are different as they are calculated from the mid-plane temperatures of the \texttt{cuDisc runs}. For the isothermal simulation, this is the same as \texttt{DustPy}, but for the non-isothermal simulation, the gas scale heights are lower due to the cooler interior of the non-isothermal disc.}
    \label{hggraindist}
\end{figure*} 

\subsection{Vertical structure comparison}

To compare the vertical structure, we converted the dust surface densities from \texttt{DustPy} into 2D volume densities by calculating the diffusion-settling equilibrium in the vertical axis for each dust grain (see \citealt{taklin02}), as might be done when taking a \texttt{DustPy} output and calculating a radiative transfer simulation. Fig. \ref{hggraindist} shows the grain size distributions at 1, 2 and 3 gas scale heights, where the gas scale height is calculated from the mid-plane temperatures. This means the scale heights are the same for the isothermal and \texttt{DustPy} runs, but lower for the non-isothermal run given the cooler interior temperature. It can now be seen that the upper cut-off grain size for the isothermal and \texttt{DustPy} runs are very similar at the mid-plane but differ at increasing height. This arises because the fragmentation limit sets the maximum grain size and the fragmentation limit is proportional to gas density. The density decreases as we move away from the mid-plane in the 2D \texttt{cuDisc} runs, leading to larger Stokes numbers and, therefore, larger turbulent velocities. This lowers the fragmentation threshold on grain size. After vertical integration, this leads to the lower cut-off grain size we see in Fig. \ref{graindist1D}. For the non-isothermal simulation, we see a similar result but with the cut-off at larger sizes due to the cooler interior. After vertical integration, this leads to the larger maximum grain size seen in Fig. \ref{graindist1D}. There are also fewer particles at high altitudes for the same reason: the cooler temperature lowers the scale height of the disc, bringing the dust closer to the mid-plane. \\

The trough in the grain size distribution arises at the particle size where turbulent motions start to become a strong source of relative velocities between particles. The location of the trough is dependent on the Reynolds number, which represents the strength of turbulent motions. In \texttt{cuDisc}, the Reynolds number varies as a function of height because it depends on the gas density and sound speed; however, in \texttt{DustPy}, it is assumed to be equal to the mid-plane value everywhere. This causes the trough location to vary as a function of height in \texttt{cuDisc}, leading to a smearing out of the feature in the vertically integrated density. \\

\begin{figure}
    \centering
    \includegraphics[width=1\columnwidth]{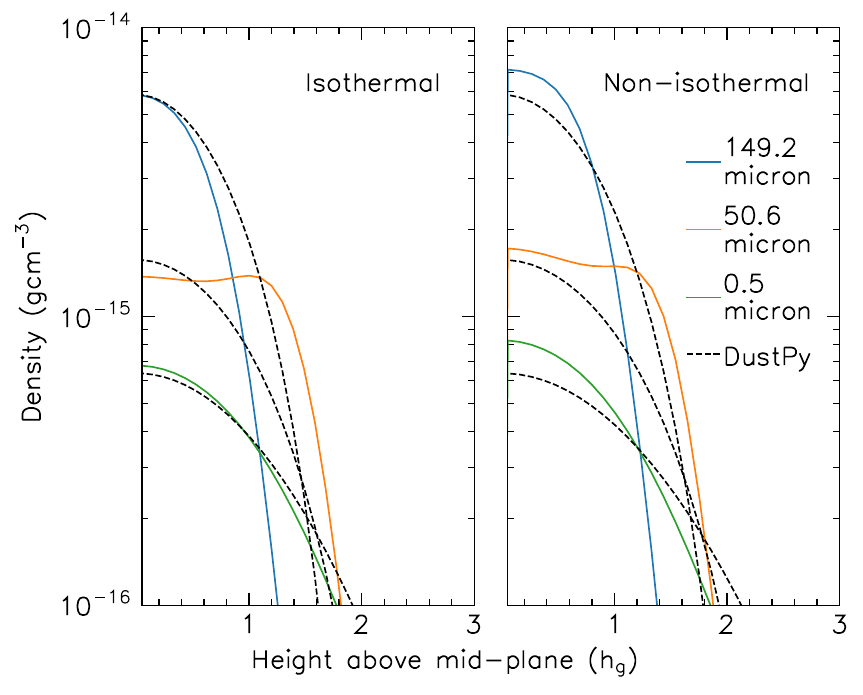}
    \caption{The vertical structure of three different grain sizes at a radius of 6 AU for the two $v_\text{frag}=100$~cm~s\minone{} \texttt{cuDisc} simulations over-plotted on the calculated \texttt{DustPy} profiles. The gas scale heights are calculated from the mid-plane temperatures of the respective \texttt{cuDisc} simulations; as the non-isothermal simulation has a cooler mid-plane, the gas scale height is lower in the right-hand panel.}
    \label{vertprofs}
\end{figure} 

Fig. \ref{vertprofs} shows the density of three different grain sizes, $\sim 0.5$, $\sim 50$ and $\sim 150$ $\mu$m, as functions of height. The seemingly lower scale height of the 150 $\mu$m grains in the \texttt{cuDisc} runs when compared to \texttt{DustPy} is due to the decrease in the maximum grain size allowed by fragmentation with height in the \texttt{cuDisc} runs discussed earlier. This decrease in maximum allowable grain size can be seen by noting that 150 $\mu$m is around the peak of the grain size distribution at the mid-plane in Fig. \ref{hggraindist} but that at 1 gas scale height, the peak has moved to $\sim$100 $\mu$m. This effect is less noticeable for the small grains (i.e. 0.5 $\mu$m in Fig. \ref{vertprofs}) as the small grain distribution is less affected by the change in the upper cut-off grain size. The non-isothermal vertical profiles are more condensed than \texttt{DustPy}, again due to the cooler interior temperatures leading to lower scale heights. \\

\begin{figure}
    \centering
    \includegraphics[width=0.9\columnwidth]{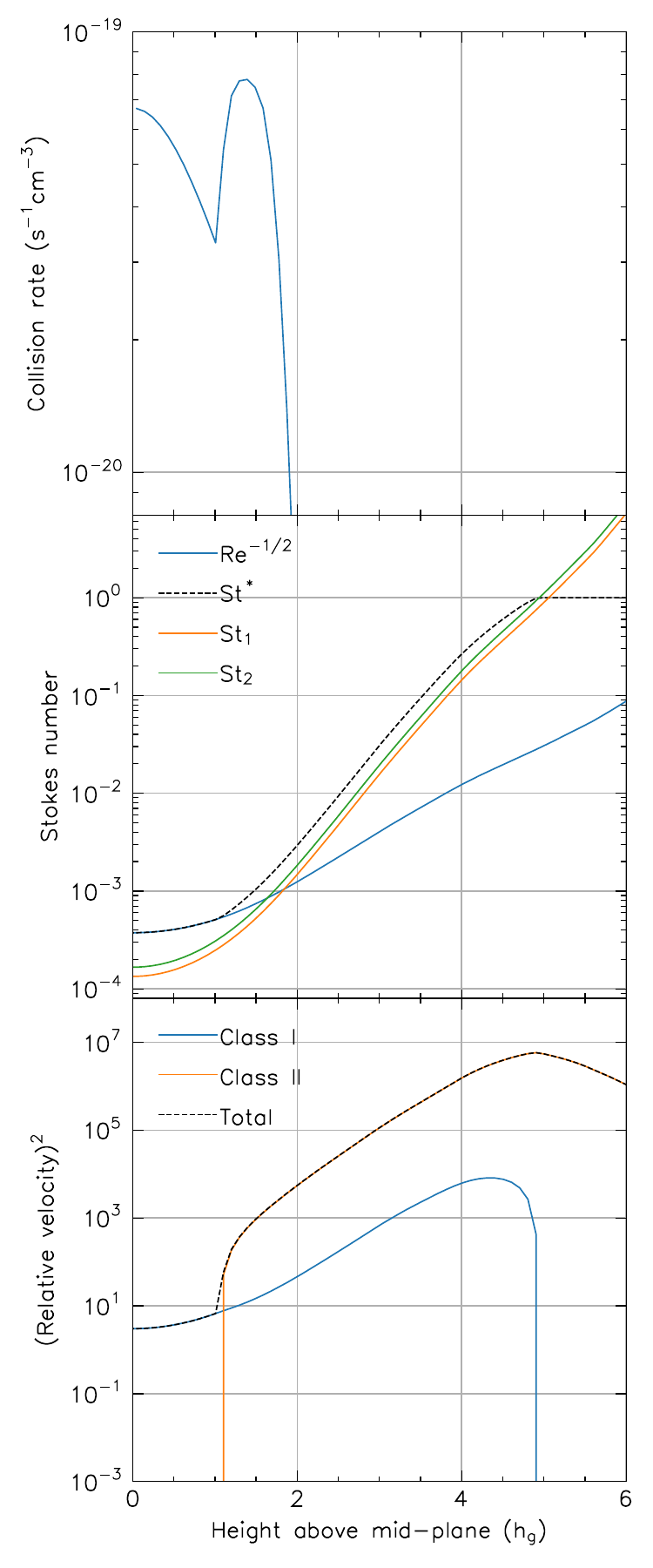}
    \caption{Collision rates, Stokes numbers and relative turbulent velocities of two different grain sizes, 19 and 24 $\mu$m, as a function of height. St$^*$ is the boundary between class I and class II eddies for a particular particle. Its minimum value is equal to the Stokes number of the smallest eddies, Re$^{-1/2}$, whilst its maximum value is equal to the Stokes number associated with the largest eddies (one orbit), 1.}
    \label{coagrates}
\end{figure}

The vertical profiles found using \texttt{cuDisc} also show increases in the amount of intermediate size (10-100 $\mu$m) grains away from the mid-plane, with some grain sizes having a higher density at around a gas scale height than at the mid-plane. To investigate this, we need to look at how the collision rates vary with height. Fig. \ref{coagrates} shows the collision rate, Stokes numbers and relative turbulent velocity of grains with sizes 19 and 24 $\mu$m as functions of height for the non-isothermal run. Re is the Reynolds number that describes the strength of turbulent viscosity over molecular viscosity, i.e. Re$ = \nu_t/\nu_\text{mol}$, whilst St$^*$ is the Stokes number for the boundary between slow and fast eddies (class I and class II in the terminology used by \citealt{ormel2007}) for a particular particle. Slow eddies have turn-over times longer than particle stopping times and induce large-scale systematic motion in the grains, whilst fast eddies have turn-over times smaller than the particle stopping times and, therefore, induce stochastic motions in the particles \citep{volk1980}. Slow eddies do not drive large relative velocities between similarly sized grains, whereas fast eddies do \citep{ormel2007}. Close to the mid-plane, the Stokes numbers of the grains are smaller than the Stokes number associated with the smallest turbulent eddies (Re$^{-1/2}$), meaning the particles only experience slow eddies; therefore, the relative velocities are low. As we move to higher regions of the disc, the particle Stokes numbers are in the intermediate regime, and particles feel the impact of fast eddies. These stochastic motions lead to larger relative velocities between similarly sized particles. The densities of the intermediate-sized grains are enhanced at these heights due to the increase in collision rates in combination with the decrease in maximum grain size with height discussed previously. Higher up in the disc, settling causes a decrease in dust density that leads to a sharp decrease in collisions regardless of large turbulent velocities, counteracting the enhancement seen at around one gas scale height.
\subsection{Particle sweeping}\label{particlesweeping}

\begin{figure*}
    \centering
    \includegraphics[width=1\textwidth]{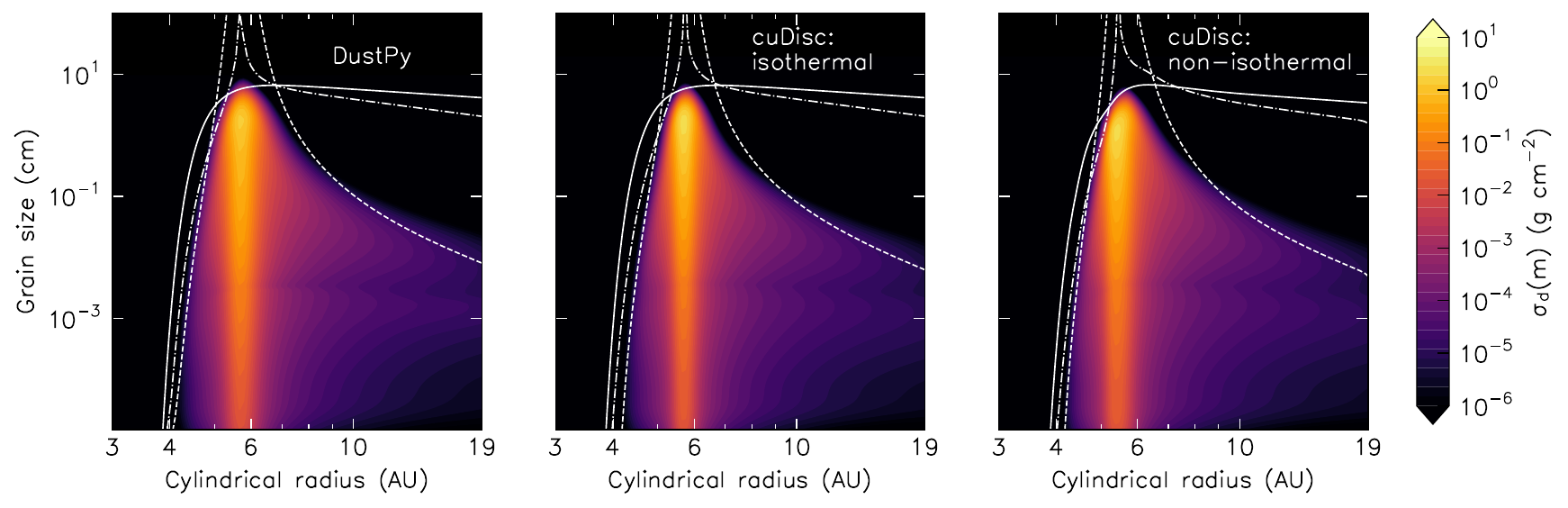}
    \caption{Dust grain size distributions as a function of disc radius for each $v_\text{frag} = 1000$~cm~s\minone{} simulation. The fragmentation and drift limits (given by \citealt{birnstiel2012}) are over-plotted for reference in solid white and dashed white respectively.}
    \label{graindist2D_1000}
\end{figure*} 

\begin{figure}
    \centering
    \includegraphics[width=1\columnwidth]{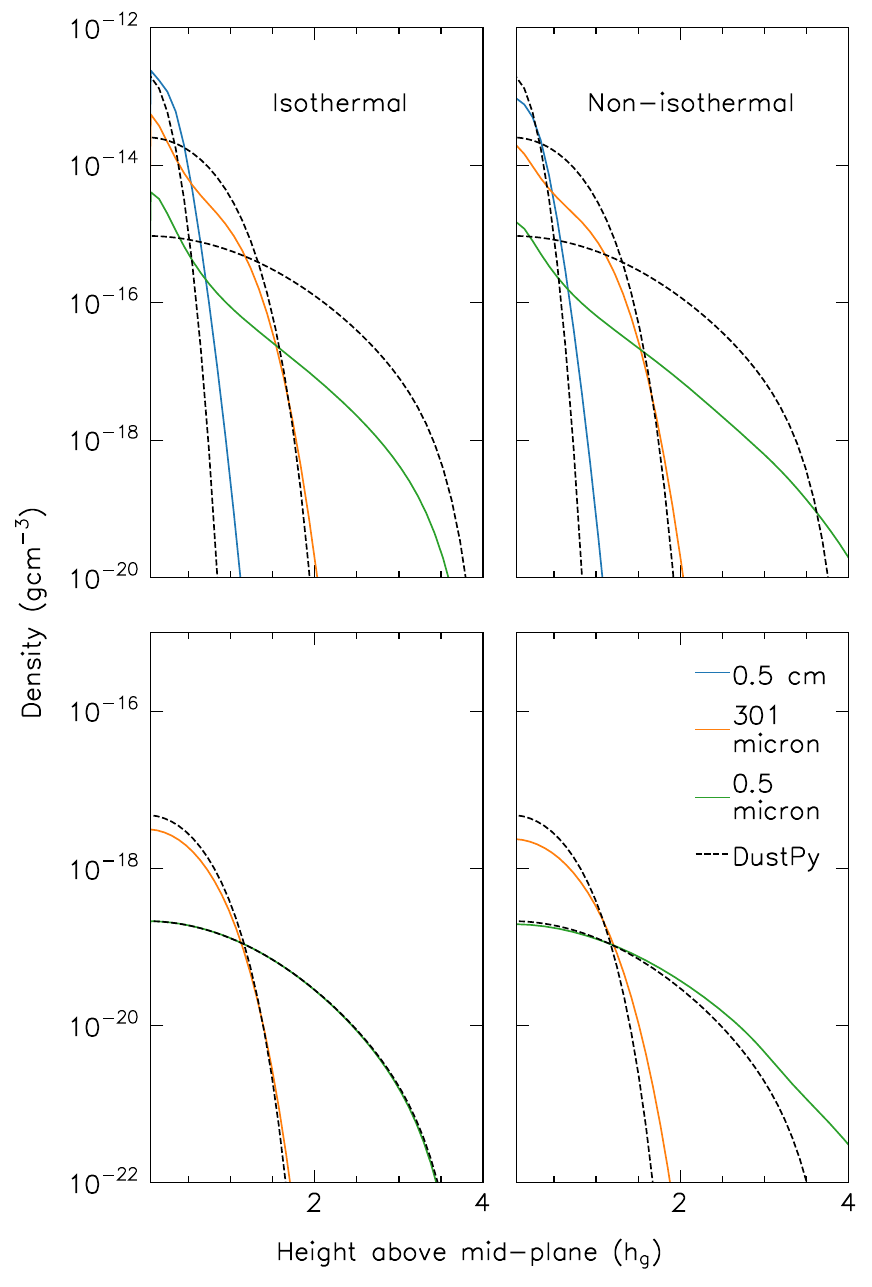}
    \caption{The vertical structure of three different grain sizes for the $v_\text{frag} = 1000$~cm~s\minone{} simulations. The top and bottom rows show the profiles within the dust trap ($\sim$5.7 AU), and away from the dust trap ($\sim 10$ AU) respectively. The density of the 0.5 cm grain is at the floor value at 10 AU in all of the simulations due to being above the drift limit. The gas scale heights are calculated from the mid-plane temperatures of the respective \texttt{cuDisc} simulations.}
    \label{vertprofs_1000}
\end{figure} 

These results differ from those seen by \cite{krijt2016}, who found that small grains become trapped in the disc mid-plane due to colliding with, and sticking to, larger grains before vertical mixing can distribute them higher in the disc. This ``sweeping-up'' of small grains by large grains decreases the abundance of small grains at high altitudes in regions where particle collision rates at the mid-plane are high compared to the rate of diffusion due to turbulent processes. A reason why we do not see this effect may be that the largest grains in our simulations are $\sim 150$ $\mu$m - these grains are less strongly settled than the cm-sized grains reached in the \cite{krijt2016} models, lessening the effect that leads to trapping in the mid-plane. To investigate this, we ran the same set of simulations with a higher fragmentation velocity of 1000~cm~s\minone{}. Fig. \ref{graindist2D_1000} shows the vertically-integrated density as a function of radius and grain size for this set of simulations, now with the addition of the approximate drift limit. Both the \texttt{cuDisc} and the \texttt{DustPy} simulations show very similar dust distributions, with the majority of the disc exhibiting drift-limited growth and the largest grains in the dust trap reaching $\sim$ a few cm. Turning to the vertical structure, Fig. \ref{vertprofs_1000} shows the density profiles of three different dust species from each simulation at the peak of the dust trap ($\sim$5.7 AU) on the top row and away from the trap ($\sim 10$ AU) on the bottom row. The \texttt{cuDisc} dust profiles in the trap clearly exhibit the trapping of smaller grains around the mid-plane, as found by \cite{krijt2016}. However, the effect is not visible at 10 AU, where the isothermal profiles closely follow the \cite{taklin02} analytic profiles calculated from the \texttt{DustPy} densities, and the non-isothermal profiles differ at a few scale heights due to the hot surface layers that increase the turbulent velocities, lofting the smaller grains to higher heights. The lack of sweeping at these radii is to be expected if the cause of the effect is the presence of large cm-sized grains, because here the the maximum grain size is lower due to the disc being in the drift-limited regime.\\

\begin{figure}
    \centering
    \includegraphics[width=.9\columnwidth, trim={.2cm 0 0cm 0}, clip]{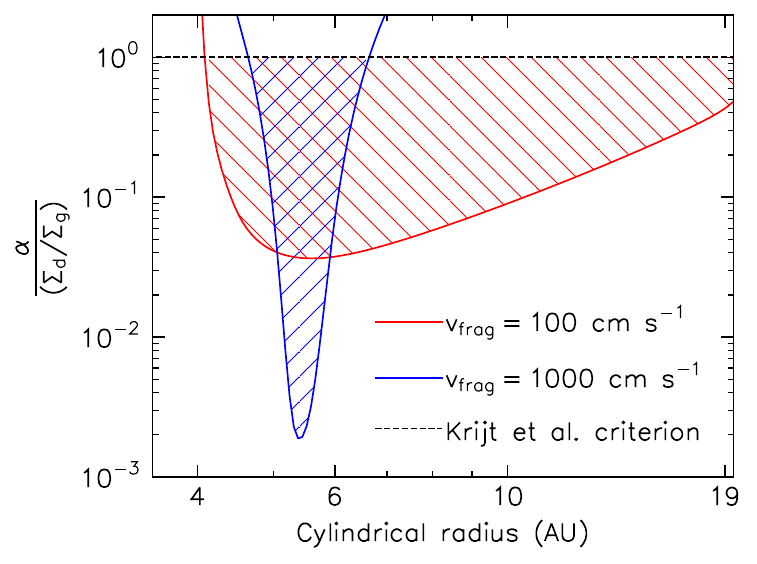}
    \caption{The regions of the disc within which the \protect\cite{krijt2016} criterion for mid-plane trapping of small grains is satisfied, shown as hatched regions, for both the low and high fragmentation velocity \texttt{cuDisc} simulations.}
    \label{sweepcrit}
\end{figure}

\cite{krijt2016} found that the effect of sweeping should be important in regions where the collisional timescale of grains is less than the timescale associated with diffusion to higher regions of the disc. In terms of disc quantities, this criterion can be written as $\alpha/(\Sigma_d/\Sigma_g) < 1$. Fig. \ref{sweepcrit} shows the radial regions of the discs in both the low and high fragmentation velocity simulations where this criterion is satisfied. In the high fragmentation velocity disc, the condition is only met around the peak of the dust trap - this concurs with our findings, as shown in Fig. \ref{vertprofs_1000}. For the low fragmentation velocity case, the condition is satisfied for the entirety of the disc; but we find no evidence of trapping. We suspect this is due to the difference in scale height between the largest and smallest grains in the simulations as discussed above; for the low fragmentation velocity simulation, the difference between the scale heights of the 0.5 and 150 micron grains is a factor of $\sim 2$, whilst, for the high fragmentation velocity simulation, the difference in scale height between the 0.5 micron and 1 cm grains is a factor of $\sim 80$. In the low fragmentation velocity case, this means that any sweeping by the largest grains does not manifest itself as an appreciable change in the vertical density distribution of the smaller grains, as the sweeping grains and swept-up grains have similar scale heights. The effect is noticeable, however, in the high fragmentation velocity case, where the largest grains are much more settled than the smallest grains.

\subsection{Diffusion-settling-coagulation equilibrium}

These results indicate that the diffusion-settling equilibrium profiles \citep{taklin02} for the dust vertical structure do not fully describe the dust population in regions where collisions are important. In these regions, we suspect that a diffusion-settling-coagulation equilibrium is established; the form of which varies depending on the fragmentation velocity. For a low fragmentation velocity of 100~cm~s\minone{}, we find enhancements of intermediate-sized dust grains at $\sim$ one gas scale height, whilst for a high fragmentation velocity of 1000~cm~s\minone{}, we find that large grains sweeping up smaller grains in the dust trap leads to the enhancement of small grains at the disc mid-plane. These findings may have implications for the analysis of observed disc spectral energy distributions (SEDs) as one must assume a disc structure to estimate the emission layers of different-sized dust grains \citep[see e.g.][]{dalessio2006}. Scattered light images could also be affected, as mid-plane densities calculated from the observed small grain distribution at large disc heights are dependent on the assumed vertical structure of said grains. However, our results also demonstrate that if one only cares about overall, general, disc evolution and not specific problems that require a resolved vertical dimension (e.g. winds or temperature instabilities caused by vertical structure), the 1D results from \texttt{DustPy} are a good approximation to the full 2D problem. 
\section{Summary}

\texttt{cuDisc} is a new protoplanetary disc code that aims to allow long time-scale calculations of discs with self-consistently calculated dynamics, thermodynamics and dust grain growth and fragmentation. Modelling these physical processes alongside one another allows us to answer important problems in disc physics, such as structure formation due to instabilities and dust removal. With the use of GPU acceleration, \texttt{cuDisc} enables simulations to run for large fractions of the disc lifetime, removing the need for running simplified secular models. We have shown that 2D structure affects the dust spatial and grain size distributions even for simple systems, such as dust trapped in the pressure bump of a transition disc. This may be important when analysing SEDs and scattered light images, given the need to assume some model of grain vertical structure. We also find that for studying overall disc evolution, 1D models can quite accurately concur with 2D models. More features can and will be added to the code as we continue to develop it, such as more detailed dust microphysics and ice-vapour chemistry, making many other problems in disc physics able to be investigated through the use of \texttt{cuDisc}.

\section*{Acknowledgements}
\HL{We would like to thank the referee for their detailed and helpful comments.} RAB and JEO are supported by Royal Society University Research Fellowships. This work has received funding from the European Research Council (ERC) under the European Union’s Horizon 2020 research and innovation programme (Grant agreement No. 853022, PEVAP) and a Royal Society Enhancement Award. Some of this work was performed using the Cambridge Service for Data Driven Discovery (CSD3), part of which is operated by the University of Cambridge Research Computing on behalf of the STFC DiRAC HPC Facility (\url{www.dirac.ac.uk}). The DiRAC component of CSD3 was funded by BEIS capital funding via STFC capital grants ST/P002307/1 and ST/R002452/1 and STFC operations grant ST/R00689X/1. DiRAC is part of the National e-Infrastructure. For the purpose of open access, the authors have applied a Creative Commons Attribution (CC-BY) licence to any Author Accepted Manuscript version arising.

\section*{Data Availability}

The GitHub repository for \texttt{cuDisc} can be found at \url{https://github.com/cuDisc/cuDisc/}.



\bibliographystyle{mnras}
\bibliography{biblio} 

\begin{thebibliography}{}
\makeatletter
\relax
\def\mn@urlcharsother{\let\do\@makeother \do\$\do\&\do\#\do\^\do\_\do\%\do\~}
\def\mn@doi{\begingroup\mn@urlcharsother \@ifnextchar [ {\mn@doi@}
  {\mn@doi@[]}}
\def\mn@doi@[#1]#2{\def\@tempa{#1}\ifx\@tempa\@empty \href
  {http://dx.doi.org/#2} {doi:#2}\else \href {http://dx.doi.org/#2} {#1}\fi
  \endgroup}
\def\mn@eprint#1#2{\mn@eprint@#1:#2::\@nil}
\def\mn@eprint@arXiv#1{\href {http://arxiv.org/abs/#1} {{\tt arXiv:#1}}}
\def\mn@eprint@dblp#1{\href {http://dblp.uni-trier.de/rec/bibtex/#1.xml}
  {dblp:#1}}
\def\mn@eprint@#1:#2:#3:#4\@nil{\def\@tempa {#1}\def\@tempb {#2}\def\@tempc
  {#3}\ifx \@tempc \@empty \let \@tempc \@tempb \let \@tempb \@tempa \fi \ifx
  \@tempb \@empty \def\@tempb {arXiv}\fi \@ifundefined
  {mn@eprint@\@tempb}{\@tempb:\@tempc}{\expandafter \expandafter \csname
  mn@eprint@\@tempb\endcsname \expandafter{\@tempc}}}

\bibitem[\protect\citeauthoryear{Andrews}{Andrews}{2020}]{andrews20}
Andrews S.~M.,  2020, \mn@doi [Annual Review of Astronomy and Astrophysics]
  {10.1146/annurev-astro-031220-010302}, 58, 483

\bibitem[\protect\citeauthoryear{{Bae}, {Isella}, {Zhu}, {Martin}, {Okuzumi}
  \& {Suriano}}{{Bae} et~al.}{2022}]{bae22}
{Bae} J.,  {Isella} A.,  {Zhu} Z.,  {Martin} R.,  {Okuzumi} S.,   {Suriano} S.,
   2022, \mn@doi [arXiv e-prints] {10.48550/arXiv.2210.13314}, \href
  {https://ui.adsabs.harvard.edu/abs/2022arXiv221013314B} {p. arXiv:2210.13314}

\bibitem[\protect\citeauthoryear{{Balbus} \& {Papaloizou}}{{Balbus} \&
  {Papaloizou}}{1999}]{balbus1999}
{Balbus} S.~A.,  {Papaloizou} J. C.~B.,  1999, \mn@doi [\apj] {10.1086/307594},
  \href {https://ui.adsabs.harvard.edu/abs/1999ApJ...521..650B} {521, 650}

\bibitem[\protect\citeauthoryear{{Batalha} et~al.,}{{Batalha}
  et~al.}{2013}]{kepler2013}
{Batalha} N.~M.,  et~al., 2013, \mn@doi [\apjs] {10.1088/0067-0049/204/2/24},
  \href {https://ui.adsabs.harvard.edu/abs/2013ApJS..204...24B} {204, 24}

\bibitem[\protect\citeauthoryear{{Birnstiel}, {Dullemond}  \&
  {Brauer}}{{Birnstiel} et~al.}{2010}]{birnstiel2010}
{Birnstiel} T.,  {Dullemond} C.~P.,   {Brauer} F.,  2010, \mn@doi [\aap]
  {10.1051/0004-6361/200913731}, \href
  {https://ui.adsabs.harvard.edu/abs/2010A&A...513A..79B} {513, A79}

\bibitem[\protect\citeauthoryear{{Birnstiel}, {Klahr}  \&
  {Ercolano}}{{Birnstiel} et~al.}{2012}]{birnstiel2012}
{Birnstiel} T.,  {Klahr} H.,   {Ercolano} B.,  2012, \mn@doi [\aap]
  {10.1051/0004-6361/201118136}, \href
  {https://ui.adsabs.harvard.edu/abs/2012A&A...539A.148B} {539, A148}

\bibitem[\protect\citeauthoryear{Birnstiel et~al.,}{Birnstiel
  et~al.}{2018}]{DSHARP2018}
Birnstiel T.,  et~al., 2018, \mn@doi [\apjl] {10.3847/2041-8213/aaf743}, 869,
  L45

\bibitem[\protect\citeauthoryear{{Bitsch}, {Crida}, {Morbidelli}, {Kley}  \&
  {Dobbs-Dixon}}{{Bitsch} et~al.}{2013}]{bitsch2013}
{Bitsch} B.,  {Crida} A.,  {Morbidelli} A.,  {Kley} W.,   {Dobbs-Dixon} I.,
  2013, \mn@doi [\aap] {10.1051/0004-6361/201220159}, \href
  {https://ui.adsabs.harvard.edu/abs/2013A&A...549A.124B} {549, A124}

\bibitem[\protect\citeauthoryear{Bogacki \& Shampine}{Bogacki \&
  Shampine}{1989}]{BOGACKI1989321}
Bogacki P.,  Shampine L.,  1989, \mn@doi [Applied Mathematics Letters]
  {https://doi.org/10.1016/0893-9659(89)90079-7}, 2, 321

\bibitem[\protect\citeauthoryear{Booth \& Clarke}{Booth \&
  Clarke}{2021}]{booth21}
Booth R.~A.,  Clarke C.~J.,  2021, \mn@doi [\mnras] {10.1093/mnras/stab090},
  502, 1569

\bibitem[\protect\citeauthoryear{Booth, Clarke, Madhusudhan  \& Ilee}{Booth
  et~al.}{2017}]{booth17}
Booth R.~A.,  Clarke C.~J.,  Madhusudhan N.,   Ilee J.~D.,  2017, \mn@doi
  [\mnras] {10.1093/mnras/stx1103}, 469, 3994

\bibitem[\protect\citeauthoryear{{Brauer}, {Dullemond}  \& {Henning}}{{Brauer}
  et~al.}{2008}]{brauer2008}
{Brauer} F.,  {Dullemond} C.~P.,   {Henning} T.,  2008, \mn@doi [\aap]
  {10.1051/0004-6361:20077759}, \href
  {https://ui.adsabs.harvard.edu/abs/2008A&A...480..859B} {480, 859}

\bibitem[\protect\citeauthoryear{{Chiang} \& {Goldreich}}{{Chiang} \&
  {Goldreich}}{1997}]{chiang1997}
{Chiang} E.~I.,  {Goldreich} P.,  1997, \mn@doi [\apj] {10.1086/304869}, \href
  {https://ui.adsabs.harvard.edu/abs/1997ApJ...490..368C} {490, 368}

\bibitem[\protect\citeauthoryear{{Clarke} \& {Pringle}}{{Clarke} \&
  {Pringle}}{1988}]{clarke1988}
{Clarke} C.~J.,  {Pringle} J.~E.,  1988, \mn@doi [\mnras]
  {10.1093/mnras/235.2.365}, \href
  {https://ui.adsabs.harvard.edu/abs/1988MNRAS.235..365C} {235, 365}

\bibitem[\protect\citeauthoryear{{Commer{\c{c}}on}, {Teyssier}, {Audit},
  {Hennebelle}  \& {Chabrier}}{{Commer{\c{c}}on} et~al.}{2011}]{Commercon2011}
{Commer{\c{c}}on} B.,  {Teyssier} R.,  {Audit} E.,  {Hennebelle} P.,
  {Chabrier} G.,  2011, \mn@doi [\aap] {10.1051/0004-6361/201015880}, \href
  {https://ui.adsabs.harvard.edu/abs/2011A&A...529A..35C} {529, A35}

\bibitem[\protect\citeauthoryear{Courant, Friedrichs  \& Lewy}{Courant
  et~al.}{1928}]{Courant_1928}
Courant R.,  Friedrichs K.,   Lewy H.,  1928, \mn@doi [Mathematische Annalen]
  {10.1007/BF01448839}, 100, 32

\bibitem[\protect\citeauthoryear{{Cuzzi}, {Dobrovolskis}  \&
  {Champney}}{{Cuzzi} et~al.}{1993}]{cuzzi1993}
{Cuzzi} J.~N.,  {Dobrovolskis} A.~R.,   {Champney} J.~M.,  1993, \mn@doi
  [\icarus] {10.1006/icar.1993.1161}, \href
  {https://ui.adsabs.harvard.edu/abs/1993Icar..106..102C} {106, 102}

\bibitem[\protect\citeauthoryear{{D'Alessio}, {Calvet}, {Hartmann},
  {Franco-Hern{\'a}ndez}  \& {Serv{\'\i}n}}{{D'Alessio}
  et~al.}{2006}]{dalessio2006}
{D'Alessio} P.,  {Calvet} N.,  {Hartmann} L.,  {Franco-Hern{\'a}ndez} R.,
  {Serv{\'\i}n} H.,  2006, \mn@doi [\apj] {10.1086/498861}, \href
  {https://ui.adsabs.harvard.edu/abs/2006ApJ...638..314D} {638, 314}

\bibitem[\protect\citeauthoryear{{Dohnanyi}}{{Dohnanyi}}{1969}]{1969JGR....74.2531D}
{Dohnanyi} J.~S.,  1969, \mn@doi [\jgr] {10.1029/JB074i010p02531}, \href
  {https://ui.adsabs.harvard.edu/abs/1969JGR....74.2531D} {74, 2531}

\bibitem[\protect\citeauthoryear{Draine}{Draine}{2003}]{Draine_2003}
Draine B.~T.,  2003, \mn@doi [\apj] {10.1086/379123}, 598, 1026

\bibitem[\protect\citeauthoryear{{Dubrulle}, {Morfill}  \&
  {Sterzik}}{{Dubrulle} et~al.}{1995}]{dubrulle1995}
{Dubrulle} B.,  {Morfill} G.,   {Sterzik} M.,  1995, \mn@doi [\icarus]
  {10.1006/icar.1995.1058}, \href
  {https://ui.adsabs.harvard.edu/abs/1995Icar..114..237D} {114, 237}

\bibitem[\protect\citeauthoryear{{Dullemond}, {Juhasz}, {Pohl}, {Sereshti},
  {Shetty}, {Peters}, {Commercon}  \& {Flock}}{{Dullemond}
  et~al.}{2012}]{2012ascl.soft02015D}
{Dullemond} C.~P.,  {Juhasz} A.,  {Pohl} A.,  {Sereshti} F.,  {Shetty} R.,
  {Peters} T.,  {Commercon} B.,   {Flock} M.,  2012, {RADMC-3D: A multi-purpose
  radiative transfer tool}, Astrophysics Source Code Library, record
  ascl:1202.015

\bibitem[\protect\citeauthoryear{Eistrup}{Eistrup}{2023}]{eistrup23}
Eistrup C.,  2023, \mn@doi [ACS Earth and Space Chemistry]
  {10.1021/acsearthspacechem.1c00342}, 7, 260

\bibitem[\protect\citeauthoryear{Ercolano \& Pascucci}{Ercolano \&
  Pascucci}{2017}]{ercolano17}
Ercolano B.,  Pascucci I.,  2017, \mn@doi [Royal Society Open Science]
  {10.1098/rsos.170114}, 4, 170114

\bibitem[\protect\citeauthoryear{{Franz}, {Picogna}, {Ercolano}  \&
  {Birnstiel}}{{Franz} et~al.}{2020}]{franz2020}
{Franz} R.,  {Picogna} G.,  {Ercolano} B.,   {Birnstiel} T.,  2020, \mn@doi
  [\aap] {10.1051/0004-6361/201936615}, \href
  {https://ui.adsabs.harvard.edu/abs/2020A&A...635A..53F} {635, A53}

\bibitem[\protect\citeauthoryear{Fritsch \& Carlson}{Fritsch \&
  Carlson}{1980}]{fritsch80}
Fritsch F.~N.,  Carlson R.~E.,  1980, \mn@doi [SIAM Journal on Numerical
  Analysis] {10.1137/0717021}, 17, 238

\bibitem[\protect\citeauthoryear{Fuksman \& Klahr}{Fuksman \&
  Klahr}{2022}]{David_Melon_Fuksman_2022}
Fuksman J. D.~M.,  Klahr H.,  2022, \mn@doi [\apj] {10.3847/1538-4357/ac7fee},
  936, 16

\bibitem[\protect\citeauthoryear{{Garaud}, {Meru}, {Galvagni}  \&
  {Olczak}}{{Garaud} et~al.}{2013}]{garaud2013}
{Garaud} P.,  {Meru} F.,  {Galvagni} M.,   {Olczak} C.,  2013, \mn@doi [\apj]
  {10.1088/0004-637X/764/2/146}, \href
  {https://ui.adsabs.harvard.edu/abs/2013ApJ...764..146G} {764, 146}

\bibitem[\protect\citeauthoryear{{Huang} \& {Bai}}{{Huang} \&
  {Bai}}{2022}]{huang2022}
{Huang} P.,  {Bai} X.-N.,  2022, \mn@doi [\apjs] {10.3847/1538-4365/ac76cb},
  \href {https://ui.adsabs.harvard.edu/abs/2022ApJS..262...11H} {262, 11}

\bibitem[\protect\citeauthoryear{{Hutchison} \& {Clarke}}{{Hutchison} \&
  {Clarke}}{2021}]{hutch2021}
{Hutchison} M.~A.,  {Clarke} C.~J.,  2021, \mn@doi [\mnras]
  {10.1093/mnras/staa3608}, \href
  {https://ui.adsabs.harvard.edu/abs/2021MNRAS.501.1127H} {501, 1127}

\bibitem[\protect\citeauthoryear{{Jankovic}, {Owen}, {Mohanty}  \&
  {Tan}}{{Jankovic} et~al.}{2021}]{jankovic2021}
{Jankovic} M.~R.,  {Owen} J.~E.,  {Mohanty} S.,   {Tan} J.~C.,  2021, \mn@doi
  [\mnras] {10.1093/mnras/stab920}, \href
  {https://ui.adsabs.harvard.edu/abs/2021MNRAS.504..280J} {504, 280}

\bibitem[\protect\citeauthoryear{{Kley}}{{Kley}}{1989}]{1989A&A...208...98K}
{Kley} W.,  1989, \aap, \href
  {https://ui.adsabs.harvard.edu/abs/1989A&A...208...98K} {208, 98}

\bibitem[\protect\citeauthoryear{{Kley}}{{Kley}}{1998}]{kley1998}
{Kley} W.,  1998, \mn@doi [\aap] {10.48550/arXiv.astro-ph/9808351}, \href
  {https://ui.adsabs.harvard.edu/abs/1998A&A...338L..37K} {338, L37}

\bibitem[\protect\citeauthoryear{{Krijt} \& {Ciesla}}{{Krijt} \&
  {Ciesla}}{2016}]{krijt2016}
{Krijt} S.,  {Ciesla} F.~J.,  2016, \mn@doi [\apj]
  {10.3847/0004-637X/822/2/111}, \href
  {https://ui.adsabs.harvard.edu/abs/2016ApJ...822..111K} {822, 111}

\bibitem[\protect\citeauthoryear{Krumholz, Ireland  \& Kratter}{Krumholz
  et~al.}{2020}]{krum2020}
Krumholz M.~R.,  Ireland M.~J.,   Kratter K.~M.,  2020, \mn@doi [\mnras]
  {10.1093/mnras/staa2546}, 498, 3023

\bibitem[\protect\citeauthoryear{{Kuiper}, {Klahr}, {Dullemond}, {Kley}  \&
  {Henning}}{{Kuiper} et~al.}{2010}]{kuiper2010}
{Kuiper} R.,  {Klahr} H.,  {Dullemond} C.,  {Kley} W.,   {Henning} T.,  2010,
  \mn@doi [\aap] {10.1051/0004-6361/200912355}, \href
  {https://ui.adsabs.harvard.edu/abs/2010A&A...511A..81K} {511, A81}

\bibitem[\protect\citeauthoryear{Leveque}{Leveque}{2004}]{leveque04}
Leveque R.~J.,  2004, \mn@doi [Journal of Hyperbolic Differential Equations]
  {10.1142/S0219891604000135}, 01, 315

\bibitem[\protect\citeauthoryear{{Levermore} \& {Pomraning}}{{Levermore} \&
  {Pomraning}}{1981}]{1981ApJ...248..321L}
{Levermore} C.~D.,  {Pomraning} G.~C.,  1981, \mn@doi [\apj] {10.1086/159157},
  \href {https://ui.adsabs.harvard.edu/abs/1981ApJ...248..321L} {248, 321}

\bibitem[\protect\citeauthoryear{Madhusudhan}{Madhusudhan}{2019}]{madhusudhan19}
Madhusudhan N.,  2019, \mn@doi [Annual Review of Astronomy and Astrophysics]
  {10.1146/annurev-astro-081817-051846}, 57, 617

\bibitem[\protect\citeauthoryear{{Mathis}, {Rumpl}  \& {Nordsieck}}{{Mathis}
  et~al.}{1977}]{MRN77}
{Mathis} J.~S.,  {Rumpl} W.,   {Nordsieck} K.~H.,  1977, \mn@doi [\apj]
  {10.1086/155591}, \href
  {https://ui.adsabs.harvard.edu/abs/1977ApJ...217..425M} {217, 425}

\bibitem[\protect\citeauthoryear{{Mayor} et~al.,}{{Mayor}
  et~al.}{2011}]{mayor2011}
{Mayor} M.,  et~al., 2011, \mn@doi [arXiv e-prints] {10.48550/arXiv.1109.2497},
  \href {https://ui.adsabs.harvard.edu/abs/2011arXiv1109.2497M} {p.
  arXiv:1109.2497}

\bibitem[\protect\citeauthoryear{{Mignone}}{{Mignone}}{2014}]{mignone2014}
{Mignone} A.,  2014, \mn@doi [Journal of Computational Physics]
  {10.1016/j.jcp.2014.04.001}, \href
  {https://ui.adsabs.harvard.edu/abs/2014JCoPh.270..784M} {270, 784}

\bibitem[\protect\citeauthoryear{Naumov}{Naumov}{2011}]{nvidiabicstab}
Naumov M.,  2011, NVIDIA White Paper

\bibitem[\protect\citeauthoryear{{{\"O}berg}, {Murray-Clay}  \&
  {Bergin}}{{{\"O}berg} et~al.}{2011}]{oberg2011}
{{\"O}berg} K.~I.,  {Murray-Clay} R.,   {Bergin} E.~A.,  2011, \mn@doi [\apjl]
  {10.1088/2041-8205/743/1/L16}, \href
  {https://ui.adsabs.harvard.edu/abs/2011ApJ...743L..16O} {743, L16}

\bibitem[\protect\citeauthoryear{{Ohtsuki}, {Nakagawa}  \&
  {Nakazawa}}{{Ohtsuki} et~al.}{1990}]{ohtsuki1990}
{Ohtsuki} K.,  {Nakagawa} Y.,   {Nakazawa} K.,  1990, \mn@doi [\icarus]
  {10.1016/0019-1035(90)90015-2}, \href
  {https://ui.adsabs.harvard.edu/abs/1990Icar...83..205O} {83, 205}

\bibitem[\protect\citeauthoryear{{Ormel} \& {Cuzzi}}{{Ormel} \&
  {Cuzzi}}{2007}]{ormel2007}
{Ormel} C.~W.,  {Cuzzi} J.~N.,  2007, \mn@doi [\aap]
  {10.1051/0004-6361:20066899}, \href
  {https://ui.adsabs.harvard.edu/abs/2007A&A...466..413O} {466, 413}

\bibitem[\protect\citeauthoryear{{Owen}}{{Owen}}{2016}]{Owen2016}
{Owen} J.~E.,  2016, \mn@doi [\pasa] {10.1017/pasa.2016.2}, \href
  {https://ui.adsabs.harvard.edu/abs/2016PASA...33....5O} {33, e005}

\bibitem[\protect\citeauthoryear{{Owen}}{{Owen}}{2020}]{owen2020}
{Owen} J.~E.,  2020, \mnras, 495, 3160

\bibitem[\protect\citeauthoryear{{Owen} \& {Kollmeier}}{{Owen} \&
  {Kollmeier}}{2019}]{owen2019}
{Owen} J.~E.,  {Kollmeier} J.~A.,  2019, \mnras, 487, 3702

\bibitem[\protect\citeauthoryear{{Pascucci}, {Wolf}, {Steinacker}, {Dullemond},
  {Henning}, {Niccolini}, {Woitke}  \& {Lopez}}{{Pascucci}
  et~al.}{2004}]{pascucci2004}
{Pascucci} I.,  {Wolf} S.,  {Steinacker} J.,  {Dullemond} C.~P.,  {Henning} T.,
   {Niccolini} G.,  {Woitke} P.,   {Lopez} B.,  2004, \mn@doi [\aap]
  {10.1051/0004-6361:20040017}, \href
  {https://ui.adsabs.harvard.edu/abs/2004A&A...417..793P} {417, 793}

\bibitem[\protect\citeauthoryear{{Philippov} \& {Rafikov}}{{Philippov} \&
  {Rafikov}}{2017}]{philippov2017}
{Philippov} A.~A.,  {Rafikov} R.~R.,  2017, \mn@doi [\apj]
  {10.3847/1538-4357/aa60ca}, \href
  {https://ui.adsabs.harvard.edu/abs/2017ApJ...837..101P} {837, 101}

\bibitem[\protect\citeauthoryear{{Pinte}, {Harries}, {Min}, {Watson},
  {Dullemond}, {Woitke}, {M{\'e}nard}  \& {Dur{\'a}n-Rojas}}{{Pinte}
  et~al.}{2009}]{pinte2009}
{Pinte} C.,  {Harries} T.~J.,  {Min} M.,  {Watson} A.~M.,  {Dullemond} C.~P.,
  {Woitke} P.,  {M{\'e}nard} F.,   {Dur{\'a}n-Rojas} M.~C.,  2009, \mn@doi
  [\aap] {10.1051/0004-6361/200811555}, \href
  {https://ui.adsabs.harvard.edu/abs/2009A&A...498..967P} {498, 967}

\bibitem[\protect\citeauthoryear{{Pringle}}{{Pringle}}{1981}]{pringle1981}
{Pringle} J.~E.,  1981, \mn@doi [\araa] {10.1146/annurev.aa.19.090181.001033},
  \href {https://ui.adsabs.harvard.edu/abs/1981ARA&A..19..137P} {19, 137}

\bibitem[\protect\citeauthoryear{{Rafikov}, {Silsbee}  \& {Booth}}{{Rafikov}
  et~al.}{2020}]{rafikov20}
{Rafikov} R.~R.,  {Silsbee} K.,   {Booth} R.~A.,  2020, \mn@doi [\apjs]
  {10.3847/1538-4365/ab7b71}, \href
  {https://ui.adsabs.harvard.edu/abs/2020ApJS..247...65R} {247, 65}

\bibitem[\protect\citeauthoryear{{Ramsey} \& {Dullemond}}{{Ramsey} \&
  {Dullemond}}{2015}]{ramsey2015}
{Ramsey} J.~P.,  {Dullemond} C.~P.,  2015, \mn@doi [\aap]
  {10.1051/0004-6361/201424954}, \href
  {https://ui.adsabs.harvard.edu/abs/2015A&A...574A..81R} {574, A81}

\bibitem[\protect\citeauthoryear{{Rodenkirch} \& {Dullemond}}{{Rodenkirch} \&
  {Dullemond}}{2022}]{roden2022}
{Rodenkirch} P.~J.,  {Dullemond} C.~P.,  2022, \mn@doi [\aap]
  {10.1051/0004-6361/202142571}, \href
  {https://ui.adsabs.harvard.edu/abs/2022A&A...659A..42R} {659, A42}

\bibitem[\protect\citeauthoryear{Roe}{Roe}{1981}]{ROE1981357}
Roe P.,  1981, \mn@doi [Journal of Computational Physics]
  {https://doi.org/10.1016/0021-9991(81)90128-5}, 43, 357

\bibitem[\protect\citeauthoryear{{Shakura} \& {Sunyaev}}{{Shakura} \&
  {Sunyaev}}{1973}]{shakira1973}
{Shakura} N.~I.,  {Sunyaev} R.~A.,  1973, \aap, 24, 337

\bibitem[\protect\citeauthoryear{{Smoluchowski}}{{Smoluchowski}}{1916}]{smol1916}
{Smoluchowski} M.~V.,  1916, Zeitschrift fur Physik, \href
  {https://ui.adsabs.harvard.edu/abs/1916ZPhy...17..557S} {17, 557}

\bibitem[\protect\citeauthoryear{Stammler \& Birnstiel}{Stammler \&
  Birnstiel}{2022}]{Stammler_2022}
Stammler S.~M.,  Birnstiel T.,  2022, \mn@doi [\apj]
  {10.3847/1538-4357/ac7d58}, 935, 35

\bibitem[\protect\citeauthoryear{Stone \& Gardiner}{Stone \&
  Gardiner}{2009}]{STONE2009139}
Stone J.~M.,  Gardiner T.,  2009, \mn@doi [New Astronomy]
  {https://doi.org/10.1016/j.newast.2008.06.003}, 14, 139

\bibitem[\protect\citeauthoryear{Takeuchi \& Lin}{Takeuchi \&
  Lin}{2002}]{taklin02}
Takeuchi T.,  Lin D. N.~C.,  2002, \mn@doi [\apj] {10.1086/344437}, 581, 1344

\bibitem[\protect\citeauthoryear{Tanaka, Inaba  \& Nakazawa}{Tanaka
  et~al.}{1996}]{TANAKA1996450}
Tanaka H.,  Inaba S.,   Nakazawa K.,  1996, \mn@doi [Icarus]
  {https://doi.org/10.1006/icar.1996.0170}, 123, 450

\bibitem[\protect\citeauthoryear{{Urpin}}{{Urpin}}{1984}]{urpin1984}
{Urpin} V.~A.,  1984, \sovast, \href
  {https://ui.adsabs.harvard.edu/abs/1984SvA....28...50U} {28, 50}

\bibitem[\protect\citeauthoryear{{V\"{o}lk}, {Jones}, {Morfill}  \&
  {Roeser}}{{V\"{o}lk} et~al.}{1980}]{volk1980}
{V\"{o}lk} H.~J.,  {Jones} F.~C.,  {Morfill} G.~E.,   {Roeser} S.,  1980, \aap,
  \href {https://ui.adsabs.harvard.edu/abs/1980A&A....85..316V} {85, 316}

\bibitem[\protect\citeauthoryear{{Watanabe} \& {Lin}}{{Watanabe} \&
  {Lin}}{2008}]{watanabe2008}
{Watanabe} S.-i.,  {Lin} D.~N.~C.,  2008, \apj, 672, 1183

\bibitem[\protect\citeauthoryear{Winn \& Fabrycky}{Winn \&
  Fabrycky}{2015}]{winn15}
Winn J.~N.,  Fabrycky D.~C.,  2015, \mn@doi [Annual Review of Astronomy and
  Astrophysics] {10.1146/annurev-astro-082214-122246}, 53, 409

\bibitem[\protect\citeauthoryear{{Woitke}, {Kamp}  \& {Thi}}{{Woitke}
  et~al.}{2009}]{woitke2009}
{Woitke} P.,  {Kamp} I.,   {Thi} W.~F.,  2009, \mn@doi [\aap]
  {10.1051/0004-6361/200911821}, \href
  {https://ui.adsabs.harvard.edu/abs/2009A&A...501..383W} {501, 383}

\bibitem[\protect\citeauthoryear{{Wootten} \& {Thompson}}{{Wootten} \&
  {Thompson}}{2009}]{alma2009}
{Wootten} A.,  {Thompson} A.~R.,  2009, \mn@doi [IEEE Proceedings]
  {10.1109/JPROC.2009.2020572}, \href
  {https://ui.adsabs.harvard.edu/abs/2009IEEEP..97.1463W} {97, 1463}

\bibitem[\protect\citeauthoryear{{Wu} \& {Lithwick}}{{Wu} \&
  {Lithwick}}{2021}]{wu2021}
{Wu} Y.,  {Lithwick} Y.,  2021, \apj, 923, 123

\bibitem[\protect\citeauthoryear{Wu, Gao  \& Dai}{Wu et~al.}{2012}]{WU20127152}
Wu J.,  Gao Z.,   Dai Z.,  2012, \mn@doi [Journal of Computational Physics]
  {https://doi.org/10.1016/j.jcp.2012.06.042}, 231, 7152

\bibitem[\protect\citeauthoryear{Zhu \& Dong}{Zhu \& Dong}{2021}]{zhu21}
Zhu W.,  Dong S.,  2021, \mn@doi [Annual Review of Astronomy and Astrophysics]
  {10.1146/annurev-astro-112420-020055}, 59, 291

\makeatother
\end{thebibliography}



\newpage
\appendix

\section{Diffusion Matrix}

The full form of $\tilde{D}_{kl}$ is found by writing out $\Tilde{F}_\sigma$ (Eqn. \ref{fluxes}) for each interface $\sigma$ about cell $i,j$ and summing all terms that apply to cell $k,l$ in the 3$\times$3 stencil of cells centred on cell $i,j$. For this exercise, we will change cell references from the form $i,j$ to the single letter form $k$, for ease of reading. As an example, referring to Fig. \ref{cellstencil}, when calculating the fluxes over the interfaces $\sigma$ and $\lambda$, $\Tilde{F}_\sigma$ and $\Tilde{F}_\lambda$ respectively, according to Eqn. \ref{F_k} we find
\begin{multline}
    \label{Fsigma}
    \Tilde{F}_\sigma = - D_k w_{k\sigma} \left[ w_{l\sigma} c_{kl} (u_l - u_k) + w_{m\tau} c_{km} (u_m - u_k) \right] + \\
    D_l w_{l\sigma} \left[ w_{k\sigma} c_{lk} (u_k - u_l) + w_{p\nu} c_{lp} (u_p - u_l) \right],
\end{multline}
\begin{multline}
    \label{Flambda}
    \Tilde{F}_\lambda = - D_a w_{a\lambda} \left[ w_{l\lambda} c_{al} (u_l - u_a) + w_{m\beta} c_{am} (u_m - u_k) \right] + \\
    D_l w_{l\lambda} \left[ w_{a\lambda} c_la (u_l - u_a) + w_{b\epsilon} c_{lb} (u_b - u_l) \right],
\end{multline}
where $\nu$, $\beta$ and $\epsilon$ are the interfaces between cells $l$ \& $p$, $a$ \& $m$ and $l$ \& $b$ respectively. \\

Taking for example cell $m$, the value in the diffusion matrix $\tilde{D}_m$ is therefore given by the sum of all terms that are multiplied by $u_m$, multiplied by their respective interface areas A, i.e.
\begin{equation}
    \label{D_m}
    \tilde{D}_m =  - D_k w_{k\sigma}w_{m\tau} c_{km} A_{\sigma} - D_a w_{a\lambda} w_{m\beta} c_{am} A_{\lambda}.
\end{equation}


\bsp	
\label{lastpage}
\end{document}